\documentclass[appendixfloats]{emulateapj-rtx4}
 \slugcomment{{\sc Accepted to ApJ:} April 24, 2015}

\usepackage{lscape}
\usepackage{amsmath}
\usepackage{epsfig}
\usepackage{graphicx}

\shorttitle{Magnetic fields of HERGs and LERGs}
\shortauthors{O'Sullivan et al.}

\begin{document}

\title{The Magnetic Field and Polarization Properties of Radio Galaxies in Different Accretion States}

\author{S.~P.~O'Sullivan\altaffilmark{1,2,3}, B.~M.~Gaensler\altaffilmark{1,3,4}, M.~A.~Lara-L\'opez\altaffilmark{2}, S.~van~Velzen\altaffilmark{5}, J.~K.~Banfield\altaffilmark{6,7}, J.~S.~Farnes\altaffilmark{1,3}}
\altaffiltext{1}{Sydney Institute for Astronomy, School of Physics, The University of Sydney, NSW 2006, Australia.}
\altaffiltext{2}{Instituto de Astronom\'ia, Universidad Nacional Aut\'onoma de M\'exico (UNAM), A.P.~70-264, 04510 M\'exico, D.F., M\'exico.}
\altaffiltext{3}{ARC Centre of Excellence for All-sky Astrophysics (CAASTRO), 44 Rosehill Street, Redfern, NSW 2016, Australia.}
\altaffiltext{4}{Dunlap Institute for Astronomy and Astrophysics, The University of Toronto, 50 St.~George Street, Toronto, ON M5S 3H4, Canada.}
\altaffiltext{5}{Department of Physics and Astronomy, The Johns Hopkins University, Baltimore, MD 21218, USA.}
\altaffiltext{6}{CSIRO Australia Telescope National Facility, PO Box 76, Epping, NSW, 1710, Australia.}
\altaffiltext{7}{Research School of Astronomy and Astrophysics, Australian National University, Weston Creek, ACT 2611, Australia.}
\email{shane@astro.unam.mx}

\begin{abstract}

We use the integrated polarized radio emission at 1.4 GHz ($\Pi_{\rm 1.4\,GHz}$) from a large sample of AGN 
(796 sources at redshifts $z<0.7$) to study the large-scale magnetic field properties of radio 
galaxies in relation to the host galaxy accretion state. We find a fundamental difference in $\Pi_{\rm 1.4\,GHz}$ 
between radiative-mode AGN (i.e.~high-excitation radio galaxies, HERGs, and radio-loud QSOs) 
and jet-mode AGN (i.e.~low-excitation radio galaxies, LERGs). While LERGs can achieve a wide 
range of $\Pi_{\rm 1.4\,GHz}$ (up to $\sim$30\%), the HERGs and radio-loud QSOs are limited to 
$\Pi_{\rm 1.4\,GHz} \lesssim 15\%$. A difference in $\Pi_{\rm 1.4\,GHz}$ is also seen when the 
sample is divided at 0.5\% of the total Eddington-scaled accretion rate, where the weakly accreting 
sources can attain higher values of $\Pi_{\rm 1.4\,GHz}$. We do not find any clear evidence that this 
is driven by intrinsic magnetic field differences of the different radio morphological classes. 
Instead, we attribute the differences in $\Pi_{\rm 1.4\,GHz}$ to the local environments of the radio 
sources, in terms of both the ambient gas density and the magnetoionic properties of this gas. 
Thus, not only are different large-scale gaseous environments potentially responsible for the different accretion 
states of HERGs and LERGs, we argue that the large-scale \emph{magnetised} environments 
may also be important for the formation of powerful AGN jets. Upcoming high angular resolution 
and broadband radio polarization surveys will provide the high precision Faraday rotation measure 
and depolarization data required to robustly test this claim.

\end{abstract}

\section{Introduction}
The interaction of powerful relativistic jets in radio-loud active galactic nuclei (AGN) 
with their environment is observed to strongly affect the dynamics of interstellar and 
intergalactic gas (e.g.~Birzan et al.~2008), likely impacting the formation and evolution 
of the most massive galaxies in the Universe (e.g.~Best et al.~2006; Croton et al.~2006). 
There exists a wide variety of radio-loud AGN jet types and powers, whose appearance is considered to 
be directly related to some combination of the intrinsic supermassive black hole properties (i.e.~mass and spin), 
the amount of material feeding the black hole, the availability of large amounts of 
magnetic flux, and the properties of the environment into which the jet propagates 
(e.g.~Stawarz et al.~2008, Laing \& Bridle~2014, Zamaninasab et al.~2014). 
One of the key open questions in the study of AGN is how the production of powerful relativistic 
jets in radio-loud AGN is related to the accretion system of the host galaxy. 
While substantial progress has been made in our understanding of the relation between the nuclear emission 
properties and the large-scale radio emission in AGN (e.g.~Rawlings \& Saunders 1991, Laing et al.~1994, 
Hardcastle et al.~2007, Buttiglione et al.~2010, de Gasperin et al.~2011, van Velzen et al.~2013, Mingo et al.~2014), 
these studies have not yet considered the magnetic properties of the source or its environment, 
even though the magnetic field is well accepted as a key parameter in AGN physics 
(e.g.~Meier 2002, Sikora \& Begelman~2013). 

The host galaxies of radio-loud AGN display a large range of optical spectral signatures that 
can be grouped into at least two main modes (Hine \& Longair 1979, Laing et al.~1994): High 
Excitation Radio Galaxies (HERGs) and Low Excitation Radio Galaxies (LERGs). The HERGs 
have strong, high-ionisation emission lines and represent the classical picture of an AGN with 
an optically-thick, geometrically-thin, radiatively-efficient accretion disk (Shakura \& Sunyaev 1973). 
This class represents the most optically-luminous AGN (including radio-loud QSOs), a population that can be 
easily identified out to high redshift ($z>3$). This class of AGN are also sometimes referred to as 
`radiative-mode', `quasar-mode' or `cold-mode' AGN. 
On the other hand, in the LERG class of objects, the optical emission 
lines are very weak or non-existent, but the presence of radio jets clearly identify the source as 
an AGN. In this case, the presence of a radiatively-inefficient accretion 
flow is inferred (e.g. Narayan \& Yi 1995). LERGs are also known as `jet-mode', `radio-mode' or `hot-mode' AGN. 
HERGs and LERGs have been proposed to be fuelled in fundamentally different 
ways (e.g.~Best et al.~2005). 
The high radiative luminosities of HERGs are considered to be fuelled by a large supply of cold gas for accretion, 
potentially provided by a merger with a gas-rich disk galaxy or internal dynamical processes in the host galaxy itself. 
The radiatively-inefficient, low-luminosity LERGs require much lower gas accretion rates, potentially 
provided by Bondi accretion of the hot X-ray halos surrounding their massive elliptical galaxy hosts 
(Hardcastle et al.~2007). 
 
Jets from both HERGs and LERGs are generally considered to be launched in similar manners 
from the vicinity of the central super-massive black hole of their host galaxy by large-scale, ordered 
magnetic fields (Blandford \& Znajek 1977, Blandford \& Payne 1982, Sikora \& Begelman 2013). 
These intrinsically magnetised jets can propagate to large ($\gg$~kpc) distances, carrying energy 
and magnetic fields into their ambient galactic and extragalactic environments.  
However, as the jets propagate through their host galaxy, the radio emission 
structure changes, presumably in part due to the type of interaction with the ambient environment. 
In general, the large scale radio morphology is typically divided into two main types (Fanaroff \& Riley 1974): 
low radio power Fanaroff-Riley Class 1 (FR1) sources where 
the inner jet and core have the brightest radio emission, and high radio 
power FR2 sources where the brightest radio emission is located 
at the jet termination `hotspots'. The traditional FR1/FR2 radio luminosity 
density divide occurs at $\sim10^{25} {\rm~W~Hz}^{-1}$ (Fanaroff \& Riley 1974), 
with a later study by Ledlow \& Owen~(1996) showing that the FR1/FR2 divide 
also depends on the host galaxy optical magnitude, with the division 
occurring at higher radio luminosities in more optically luminous hosts. 
This result indicates that the FR1/FR2 morphology is most likely due to a 
combination of jet power and environment, since it is thought that the 
Ledlow \& Owen~(1996) division arises due to jets being more easily disrupted 
into FR1 morphologies in the more massive host galaxies. 

The FR1 and FR2 populations also appear to differ in their accretion 
states, with HERGs often associated with FR2s and LERGs mainly 
of the FR1 type (e.g.~Laing et al.~1994). 
The common interpretation of this association is that with a higher 
accretion rate, due to the plentiful supply of cold gas, the HERGs can 
produce more powerful jets, that are more likely to develop into 
FR2-type sources; while the weakly accreting LERGs produce the 
less powerful FR1 jets. 
However, at low redshift, where both LERGs and HERGs can be equally well 
identified, there are several examples of FR1-HERGs and FR2-LERGs 
(Hardcastle et al.~2007, Best~2009, Buttiglione et al.~2010).
A detailed comparison of the 
FR-type and host galaxy accretion mode found that the FR class 
was independent of the HERG/LERG divide (Gendre et al.~2013). 
Thus, the nature of the relationship between the accretion mode of 
the host galaxy, the formation of powerful radio jets, the morphological 
appearance of these jets on large scales, and the large scale 
environment is still unclear.

The radio polarization properties of radio-loud AGN are important because 
they probe the internal structure of the jet magnetic field,
as well as providing a unique diagnostic of the magnetoionic environment of the radio source 
through Faraday rotation of the linear polarization 
(e.g.~Laing 1988, Garrington et al.~1988, Laing et al.~2008).  
The relationship between the large scale polarization properties 
of the radio jets/lobes and the host galaxy accretion state 
is important not only because it is a sensitive probe of the local environment, 
and thus the origin of the gas fuelling the black hole, but it also probes the magnetic 
field properties of the gas, 
and therefore helps provide an assessment of the ability of the black hole to accumulate 
the sufficiently large amounts of magnetic flux required for launching powerful 
radio-loud jets (e.g.~Tchekhovskoy et al.~2012). 

In this paper, we present an analysis of the accretion properties of host galaxies (as probed 
by nuclear optical emission lines) in relation to the large-scale magnetic field properties of 
their AGN jet (as probed by linearly-polarized radio emission).  
In Section 2, we describe the construction of our sample of polarized HERG and LERG 
sources, the classification of their total intensity radio morphologies and the 
measurement of their linear sizes. 
In Section 3, we present our results on the integrated 1.4~GHz polarization properties 
of HERGs and LERGs. We also investigate the dependence of the polarization on 
radio morphology, linear size, Faraday rotation and the Eddington-scaled accretion rate. 
In Section 4, we discuss the implications of our results on the link between the large 
scale magnetised environment of HERGs and LERGs and their black hole accretion state. 
Our conclusions are listed in Section 5. 
Throughout this paper, we assume a flat $\Lambda$CDM cosmology with 
H$_0 = 67.3$ km s$^{-1}$ Mpc$^{-1}$, 
$\Omega_M=0.315$ and $\Omega_{\Lambda}=0.685$ (Planck Collaboration et al.~2014).

\section{Sample construction}

The only high-resolution radio survey covering a significant fraction of the sky ($\sim$82\%) with 
polarization sensitivity is the NRAO VLA Sky Survey (NVSS), conducted at 1.4~GHz (Condon et al.~1998). 
The NVSS has an angular resolution of 45" with an approximately uniform rms sensitivity of 
$\sim$0.45~mJy~beam$^{-1}$ in Stokes $I$ and $\sim$0.29~mJy~beam$^{-1}$ in Stokes $Q$ and $U$. 
While the NVSS catalog already lists the basic integrated polarization properties, we have generated 
our own catalog by downloading the full NVSS $IQU$ mosaics\footnote{http://www.cv.nrao.edu/nvss/} and 
using the source-finding program {\small AEGEAN} (Hancock et al.~2012). 
This was done as part of a larger project in order to better characterise the morphologies and multi-component 
nature of polarized sources in the NVSS and has generated catalog 
positions for components in polarized intensity, Stokes $Q$ and Stokes $U$, independent of Stokes $I$. 
{\small AEGEAN} is an optimised source-finder designed for reliably identifying 
compact radio sources based on a Laplacian kernel. It finds peaks in an image based on a `seed' threshold 
($\sigma_s$) and grows an island of pixels around that peak based on a `flood' threshold ($\sigma_f$), 
where the local image noise level ($\sigma_{\rm rms}$) is estimated over a $20\times20$ beam area. 
A curvature image is then used to determine how many Gaussian components are used to describe 
the island, returning a catalog of component flux densities and angular sizes. For our polarization 
source finding in the NVSS polarized intensity images ($p=(Q^2+U^2)^{1/2}$), we used $\sigma_f=7\sigma_{\rm rms}$ 
and $\sigma_s=8\sigma_{\rm rms}$, with $\sigma_{\rm rms}\sim0.2$~mJy being typical for most fields. 
This has resulted in the inclusion of only those polarized sources with peak polarized intensities 
above $8\sigma$ of the local noise level in polarized intensity. 
Polarization bias was corrected for using the estimator $p_0=(p^2 - \sigma_{QU}^2)^{1/2}$, 
where $p_0$ is an estimate of the true polarized intensity (Simmons \& Stewart 1985).  
This has produced catalogs of $\sim$87,000 components in polarized intensity 
and total intensity (Stokes $I$), as well as Stokes $Q$ and $U$ individually, over the entire NVSS 
area. A subset of that catalog is used here. The full polarization catalog will be presented elsewhere.

\subsection{HERG/LERG parent sample}
In order to calculate physically interesting quantities, we need to match each polarized radio source 
with its spectroscopically identified optical host galaxy. The largest optical spectroscopic 
survey overlapping with the NVSS is the Sloan Digital Sky Survey (SDSS), which covers $\sim$10,000 square 
degrees of the Northern Hemisphere sky (DR 7; Abazajian et al.~2009). In addition to this, 
the Faint Images of the Radio Sky at Twenty centimetres (FIRST) survey (Becker, White \& Helfand 1995), 
which overlaps with the SDSS survey area, has a similar sensitivity limit as the NVSS but a significantly 
higher angular resolution of 5", that aids substantially in obtaining reliable cross-matches with the SDSS. 
Combining the NVSS, FIRST and SDSS catalogs, Best \& Heckman~(2012), hereafter BH12, produced a 
large sample of radio-loud AGN, with spectroscopic classification 
into HERGs and LERGs out to a redshift $z\sim0.7$. They used the emission line information from 
the value-added spectroscopic catalogs available at http://www.mpa-garching.mpg.de/SDSS/ 
(cf.~Brinchmann et al.~2004) to robustly classify 481 HERGs and 9,863 LERGs out of 18,286 
radio-loud AGN. We also use the line emission information from these catalogs in this paper.

\subsection{HERG/LERG polarization sample}
We constructed the sample used in this paper by cross-matching our NVSS 
polarization catalog with the BH12 catalog of HERGs and LERGs. Since BH12 
have already carefully cross-matched the NVSS sources with their SDSS optical 
host galaxies using the additional, higher-resolution information provided by 
FIRST, our catalog is immediately highly reliable. 
Our initial catalog used a cross-matching radius of 1 arcminute and produced 802 matches. 
After manual visual inspection of the optical (SDSS) and radio (NVSS \& FIRST) images of all 
matches, we found only 6 matches that had either misidentified optical hosts or 
were in confused fields (i.e.~there was a potential contribution to $\Pi_{\rm 1.4\,GHz}$ 
by radio emission from an unrelated radio source). Thus, the final catalog has 796 polarized radio sources, 
the vast majority of which are unresolved in the NVSS ($\sim$96\%).  

In the case of multiple NVSS components related to one radio source, we first integrated 
the contribution of Stokes $Q$ and $U$ from each component before calculation of the 
polarized intensity, and then divided by the sum of the total intensity components to get 
the integrated degree of polarization, $\Pi_{\rm 1.4\,GHz}$. 
The polarization bias correction was applied to the integrated polarized intensity. 
Ideally the bias correction should be applied to each component first, but that would have 
prevented us from comparing the polarization properties of the resolved sources with the 
unresolved sources in our analysis. 
In total, there are 32 resolved sources (in Stokes $I$) in our sample. Only 3 of these resolved 
sources have multiple polarized components, 19 have polarized component major axes 
greater than 45", while the remainder have compact polarized components (major axis $\leq$~45"). 
For the 3 multiple-component polarized sources, the higher noise level of $\sigma_{QU}$ for the bias correction is 
calculated as the quadrature sum of the $\sigma_Q$ and $\sigma_U$ noise for each component.

\begin{figure}
\centering
    \includegraphics[width=7.0cm]{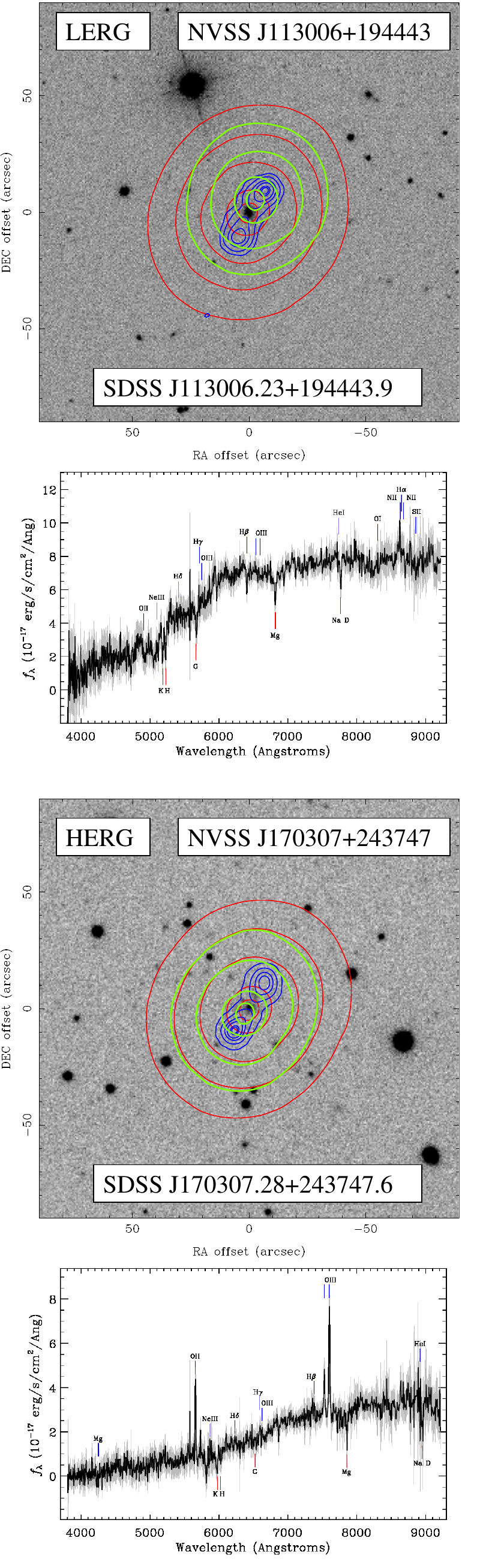}
      \caption{Example images from the cross-matched datasets, with the corresponding SDSS spectrum of
       the host galaxy included (obtained from the SDSS Science Archive Server, http://dr10.sdss3.org/).   
      Top: Low Excitation Radio Galaxy (LERG) at (RA, Dec) = (172.52602, 19.74556). 
      Bottom: High Excitation Radio Galaxy (HERG) at (RA, Dec) = (255.78027, 24.62987). 
      SDSS i-band (greyscale), NVSS total intensity (red contours), NVSS polarized intensity (green contours), 
      FIRST total intensity (blue contours). 
}
  \label{herglerg}
\end{figure}

Figure~\ref{herglerg} shows examples of two SDSS galaxies (one HERG and one LERG, 
with their accompanying optical spectrum shown) overlaid 
with the corresponding NVSS $I$ and $p$ images, as well as the FIRST total intensity image. 
Note how the radio images appear almost identical but the optical spectra differ considerably. 

\subsection{FIRST morphological analysis}
The high angular resolution FIRST images allow us to classify the total intensity morphology 
of all the polarized sources, and to calculate the projected linear size from the total angular extent 
of the FIRST total intensity emission. 
We manually classified the radio morphology of the 796 polarized sources into FR1 and FR2 
based on whether the source was edge-brightened with distinct hotspots (FR2) or edge-darkened 
with the highest surface brightness features along the jet and/or the core (FR1).  
To ensure that our results are not influenced by misclassifications, we denote source 
morphologies as `FR1?' or `FR2?' in the cases where the FIRST image 
fidelity is not of sufficient quality to conclusively decide whether it really is an FR1 or FR2.
We assign an FR0 classification (e.g.~Sadler et al.~2014) to those sources that are unresolved at 5". 
This corresponds to a projected linear extent of $\sim$15~kpc at the median redshift of our sample 
($z_{\rm median}=0.2$). Thus, these sources are most likely compact symmetric objects (CSOs), 
blazars or sources with weak jets (e.g.~low-luminosity AGN). 
The sources classified as `FR0?' are ones that have a bright core but for which it is unclear if there is 
extended emission that has been resolved out or is too faint to be detected in FIRST. 
In addition to the FR0/FR1/FR2 classification, we also manually classified the radio morphology into 
`straight', `bent' and `compact'. Sources were determined to be `straight' or `bent' depending on 
whether the extended emission on either side of the host galaxy was co-linear or not. 
The criteria for the `compact' classification is identical to the FR0 classification. 
To calculate the projected linear size of the source in kpc, we use a manual estimate 
of the angular size of each source along with the redshift from the SDSS. When the angular size of 
a source was larger than the synthesised beam of the FIRST image, we estimated the full 
angular extent of the emission by drawing a straight line between the extremities of the emission.  
Unresolved sources were assigned an angular size of 5", which corresponds to an upper 
limit on the linear size of the source. 

\subsection{Infrared (IR) data}
We also cross-matched our sample of polarized HERGs and LERGs with the 
Wide-field Infrared Survey Explorer (WISE; Wright et al. 2010), conducted 
at four bands, 3.4, 4.6, 12 and 22~$\mu$m, with $5\sigma$ point source sensitivities 
in unconfused regions of $\sim$0.08, 0.11, 1.0, and 6.0~mJy, and 
angular resolutions of 6.1, 6.4, 6.5 and 12 arcsec, in each band respectively. The IR properties across 
the WISE band are useful in determining the host galaxy types of our sample 
(e.g.~quiescent, star-forming). We found 456 reliable cross-matches out of 796 sources 
down to a $5\sigma$ detection level in at least one of the four WISE bands.
For conversion from the calibrated WISE VEGA magnitudes to flux density (in Jy), we followed the instructions on the 
WISE webpage,\footnote{http://wise2.ipac.caltech.edu/docs/release/allsky/expsup/} 
including the 10\% flux correction in the 22~$\mu$m band.


\section{Results}
The integrated degree of polarization measured at 1.4~GHz ($\Pi_{\rm 1.4\,GHz}$) 
in the NVSS can be influenced by a number of effects. 
Firstly, bright and resolved structures in radio galaxies (i.e.~knots, filaments, hotspots) 
typically do not achieve degrees of polarization larger than $\sim$30\% to 50\% (Saikia \& Salter 1988). 
This implies at least some amount of intrinsic disorder to the magnetic field on small scales 
in order to reduce it from the theoretical maximum of $\sim$72\% expected for optically thin 
synchrotron radiation in a completely uniform magnetic field 
(e.g.~Pacholczyk~1970).\footnote{Note that a completely disordered field that is compressed 
can also produce a high degree of polarization (e.g.~Laing~1981). }
Secondly, even with high intrinsic degrees of polarization on small scales, if the 
magnetic field structure is not globally uniform then polarization angle cancellation 
across the jet and lobe structure will significantly reduce the integrated degree of polarization. 
Thirdly, polarization measurements at 1.4 GHz are expected to be strongly affected by 
Faraday rotation, since the rotation of the polarization angle depends on wavelength-squared. 
Large spatial fluctuations in Faraday rotation are observed in radio galaxies (e.g.~Laing et al.~2008), 
that cause significant depolarization at 1.4~GHz. 
Finally, even if the intrinsic magnetic field is globally uniform and the Faraday rotation 
is negligible, observed asymmetries of the jet and lobe emission structure, due to physical 
bends or relativistic effects, will lead to polarization angle cancellation that can reduce the 
degree of polarization when integrated over the entire source. 

For the 796 polarized sources, we find that $\Pi_{\rm 1.4\,GHz}$ ranges 
from $>0.1\%$ to $\lesssim30\%$ (Fig.~\ref{sint_I_p}), with a median value of 6.2\%. 
This is consistent with the maximum integrated polarization expected at 1.4~GHz of 
$\lesssim30\%$ from surveys of extragalactic radio sources (e.g.~Hales et al.~2014, 
although a small number of exceptions were presented by Shi et al.~2010). 
Since our only available polarization measurements are at a single frequency 
and they have an angular resolution of 45" 
(corresponding to $\sim$150~kpc at the median redshift of 0.2 of 
the sample), we cannot study the depolarization characteristics of our 
sample or the polarization morphology in any detail. Thus, we attempt 
to gain insight into the nature of HERG and LERG sources through their 
integrated degree of polarization properties at 1.4~GHz, in conjunction with 
their optical spectroscopic properties and total intensity radio morphology.

\subsection{Integrated Degree of Polarization of HERGs and LERGs}

\begin{figure}
\centering
\vspace{-0.5cm}
    \includegraphics[width=8.5cm]{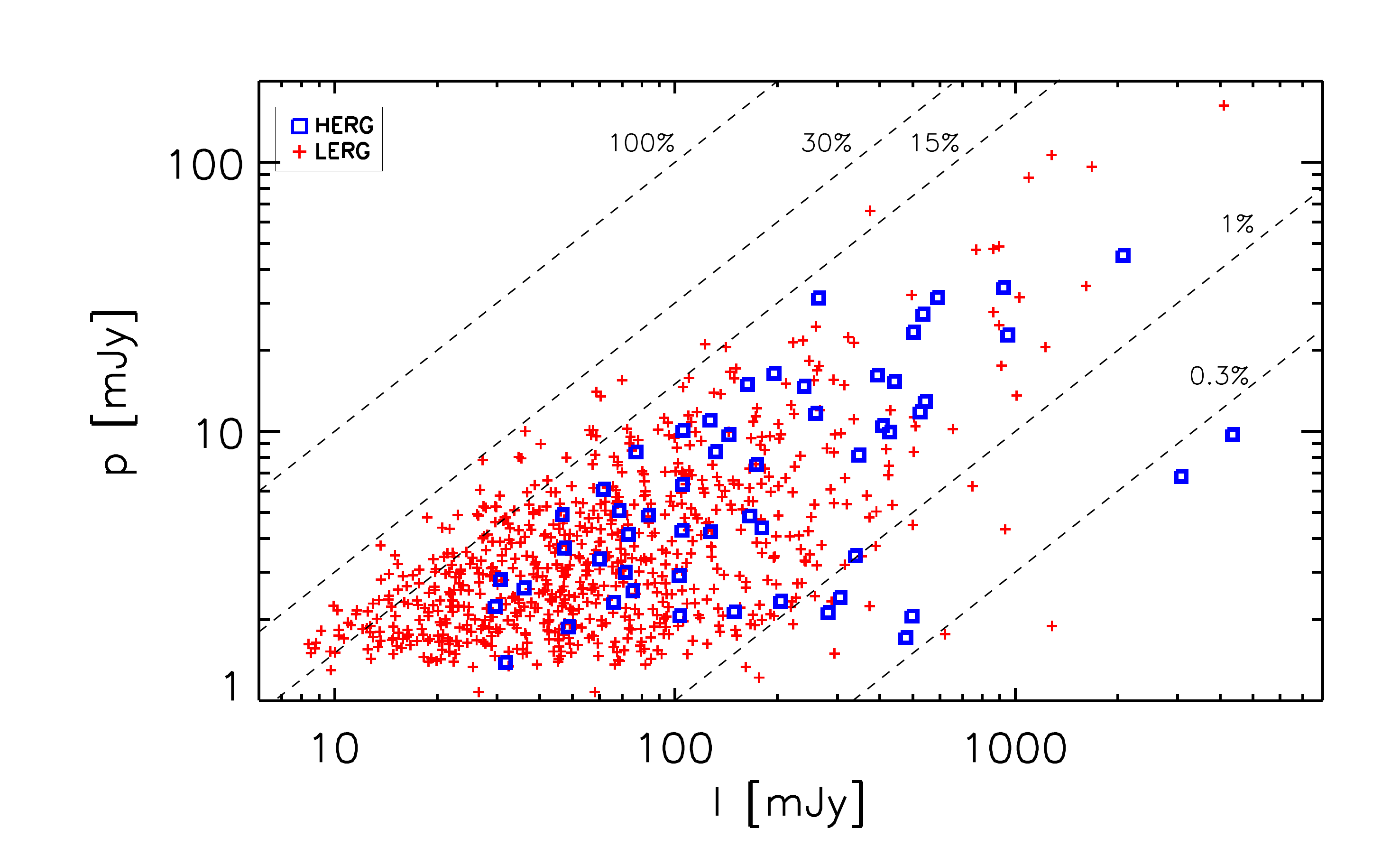}
      \caption{Integrated linear polarized flux density ($p$) versus total flux density ($I$) for HERGs (blue squares) 
      and LERGs (red plus symbols). Diagonal dashed lines represent constant values of integrated 
      degree of polarization.  
}
  \label{sint_I_p}
\end{figure}

Out of the 9,863 LERGs and 481 HERGs identified by BH12 out to $z\sim0.7$, we 
detect 741 LERGs and 55 HERGs with integrated linear polarized intensity at 1.4 GHz 
greater than $8\sigma$ ($\sim$1.6~mJy). 
There are 67 LERGs and 2 HERGs that have an integrated polarized flux density less than 1.6~mJy. 
These are still $8\sigma$ detections but have lower local noise estimates than are typical 
for the entire NVSS. 
The faintest total intensity source we detect in polarization has a Stokes $I$ flux density of 8.6~mJy. At this flux 
level, there are 393 HERGs and 7,661 LERGs meaning that $\sim$$90\%$ of sources down to this 
flux level remain undetected in polarization. 
The integrated total and polarized flux densities of the detected polarized sources are 
shown in Figure~\ref{sint_I_p} with HERGs identified as blue square symbols and LERGs as red plus symbols; 
the dashed lines are used to highlight that the majority of polarized sources have integrated 
1.4 GHz percentage polarizations ($\Pi_{\rm 1.4\,GHz}$) between 1 and 30\%. 
Both types of sources cover similar ranges in polarized intensity 
and the high fractional polarization sources are not all clustered 
at the detection limit (Fig.~\ref{sint_I_p}).  
The rms residual instrumental polarization of the NVSS is estimated as $\sim$$0.3\%$ 
(Condon et al.~1998, Stil et al.~2014). Two LERGs and two HERGs 
have $\Pi_{\rm 1.4\,GHz}<0.3\%$ (Fig.~2), and we exclude them from our subsequent analysis. 
Thus, the remaining sample is composed of 53 HERGs and 739 LERGs.

The LERG sources span the full range of expected $\Pi_{\rm 1.4\,GHz}$ from $>$$0.3\%$ to $\sim$$30\%$, 
while the detected HERGs are restricted to $\Pi_{\rm 1.4\,GHz} < 15\%$ (Fig.~\ref{hist_fpol}). 
The median $\Pi_{\rm 1.4\,GHz}$ for LERGs is 6.2\% and 4.2\% for HERGs.  
A two-sided Kolmogorov-Smirnov (KS) test of $\Pi_{\rm 1.4\,GHz}$ for HERGs and LERGs 
gives a probability or `p-value' of 0.4\%, indicating a significant rejection, at approximately $2.9\sigma$, 
of the null hypothesis (i.e.~the two datasets are not likely drawn from the same underlying 
distribution).\footnote{Throughout this paper we use the two-sample KS test to determine the probability 
(or `p-value') that the null hypothesis can be rejected for two samples being tested. In all cases, the 
null hypothesis is that the two samples are drawn from the same underlying distribution. 
In addition to the p-value, we also quote the significance level in terms of the equivalent result from a 
normally distributed process. }
In the local universe, LERGs dominate over HERGs at low radio luminosity (BH12), meaning 
that LERGs are more numerous than HERGs at low radio flux density in our volume limited sample (Fig.~\ref{sint_I_p}). 
Ricean bias effects the measurement of polarized intensity more strongly at low 
flux density (see Section 2). 
Therefore, one might expect the LERGs to be more strongly affected by 
the polarization bias than the HERGs. If we restrict our sample to those sources with polarized intensity 
greater than $8\sigma_{QU}$, then we effectively eliminate the effect of the polarization bias 
(e.g.~Simmons \& Stewart 1985).  
We also remove the 32 extended sources since the bias correction applied to them is 
not ideal for resolved sources. 
This reduces the sample to 43 HERGs and 474 LERGs, with a 
KS test of $\Pi_{\rm 1.4\,GHz}$ for these HERGs and LERGs now giving a 
p-value of 0.2\% ($\sim$3.1$\sigma$). 
This small increase in significance shows that the 
polarization bias does not strongly affect our full sample. 
 
A more important bias is the selection bias introduced by the threshold in polarized intensity at $\sim$1.6~mJy (see Fig.~2). This causes the median $\Pi_{\rm 1.4\,GHz}$  to increase for fainter sources because only sources with high values of $\Pi_{\rm 1.4\,GHz}$ will be detected. If we consider the sources for which this threshold in polarized intensity is not as important (e.g.~for sources brighter than 100~mJy), then we find no statistically significant difference between $\Pi_{\rm 1.4\,GHz}$  for HERGs and LERGs (see Appendix A). 
More sensitive polarization observations are required (by at least an order of magnitude) to conclusively test for a significant difference in $\Pi_{\rm 1.4\,GHz}$ between HERGs and LERGs down to $I\sim10$~mJy.
However, Stil et al.~(2014) circumvented this problem somewhat by stacking NVSS sources in polarized intensity to obtain the median degree of polarization for sources that were too faint to be detected. They found that the median degree of polarization increases with decreasing flux density ($\Pi_{\rm 1.4\,GHz}  \propto S_{\rm 1.4\,GHz}^{-0.051}$, where $S_{\rm 1.4\,GHz}$ is the NVSS flux density at 1.4~GHz). Our results suggest that one of the most likely causes of this increase in $\Pi_{\rm 1.4\,GHz}$ is the large number of LERGs at low flux with  $\Pi_{\rm 1.4\,GHz} > 15\%$. 
Thus, the main focus of this paper is on the absence of HERGs with  $\Pi_{\rm 1.4\,GHz} > 15\%$ and the ability of some LERGs to achieve values of $\Pi_{\rm 1.4\,GHz}$ up to $\sim$30\%.

\subsubsection{Integrated degree of polarization of radio-loud QSOs}
\begin{figure}
\centering
\vspace{-0.5cm}
    \includegraphics[width=8.5cm]{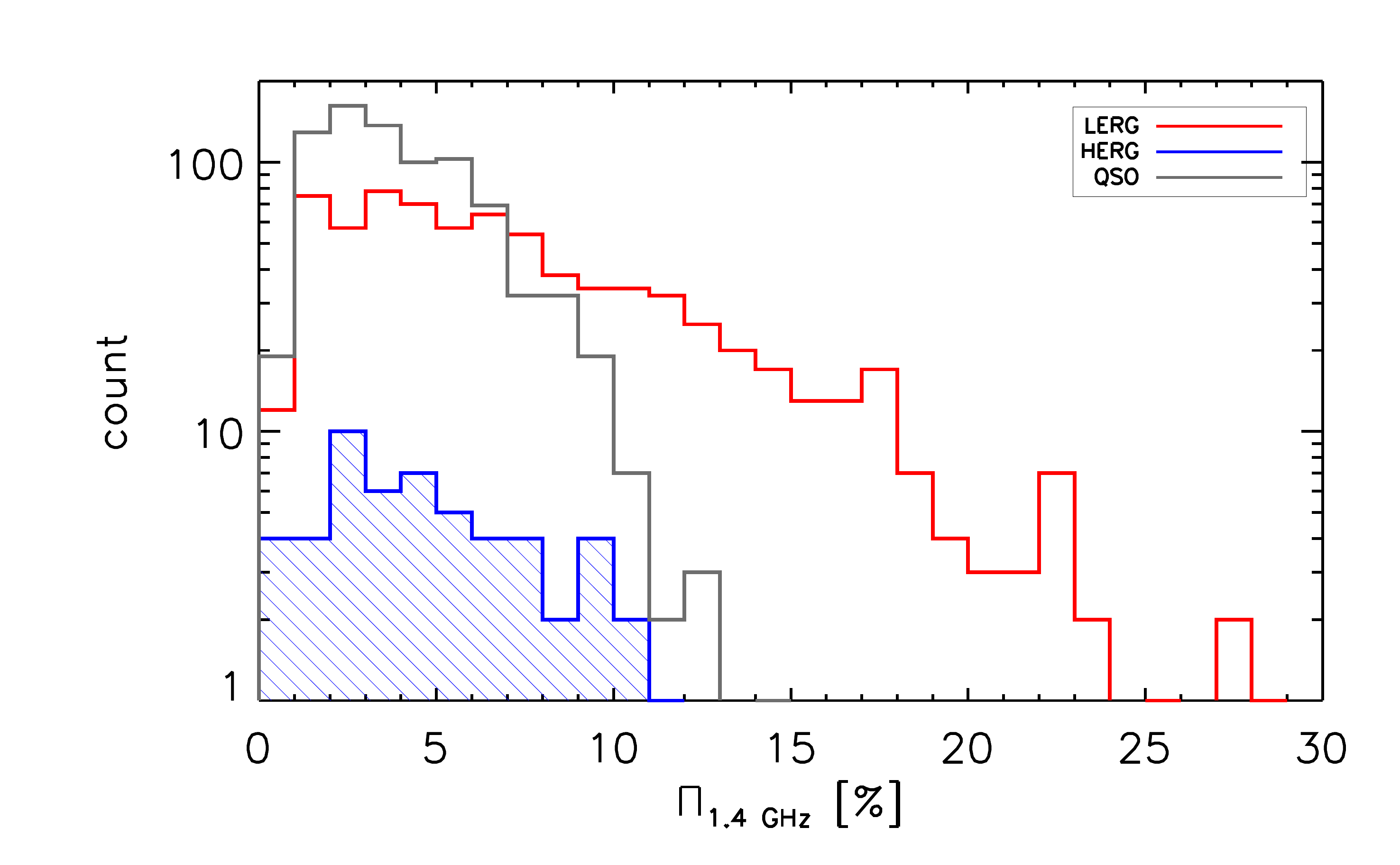}
      \caption{Histograms of the integrated degree of polarization ($\Pi_{\rm 1.4\,GHz}$) 
      for LERGs (red), HERGs (blue, hatched) and radio-loud QSOs (grey).  
}
  \label{hist_fpol}
\end{figure}

Hammond et al.~(2012) presented a catalog of 4,003 polarized radio 
sources with redshifts, 815 of which were identified as QSOs in the SDSS, 
with 89 at $z<0.5$. 
Radio-loud QSOs are the equivalent of HERGs (ie.~radio-loud AGN 
with a radiatively-efficient accretion disk) but with a more direct line of 
sight to the AGN nucleus and typically more luminous in the optical, allowing them to be 
more easily detected out to high redshift (e.g.~van Velzen et al.~2015). 
A clear difference between the 1.4~GHz fractional polarization distributions of SDSS galaxies and of 
QSOs was found by Hammond et al.~(2012), with the SDSS QSOs not exceeding fractional polarizations 
greater than $\sim15\%$ while the SDSS galaxies could reach as high as $\sim30\%$. 
Here we have essentially expanded on their sample of SDSS galaxies and separated them 
into the more physically meaningful classes of HERGs and LERGs. 
In Figure~\ref{hist_fpol}, separate histograms of $\Pi_{\rm 1.4\,GHz}$ are shown 
for the LERGs, HERGs and QSOs. 
Now there are roughly equal numbers of radiative-mode AGN (i.e.~HERGs and QSOs) and 
jet-mode AGN (i.e.~LERGs), and the maximum degree of polarization of the radiative-mode 
AGN is still limited to $\lesssim15\%$. 
This strongly suggests a fundamental difference between the 
polarization properties of radiative-mode AGN and jet-mode AGN at 1.4~GHz. 
We note that there are potentially strong redshift and evolutionary effects for the 
radio-loud QSO sample ($0.06<z<5.3$). 
Thus, we limit most of our detailed analysis and discussion to the 
polarized HERGs and LERGs because they are similarly distributed in redshift, 
with $z_{\rm median, HERG}=0.25$ and $z_{\rm median, LERG}=0.18$ (Figure~\ref{fpol_z}). 
In Appendix~A, we replot Figure~\ref{fpol_z} for the sources with Stokes $I>100$~mJy only. 

\begin{figure}
\centering
\vspace{-0.5cm}
\includegraphics[width=8.5cm]{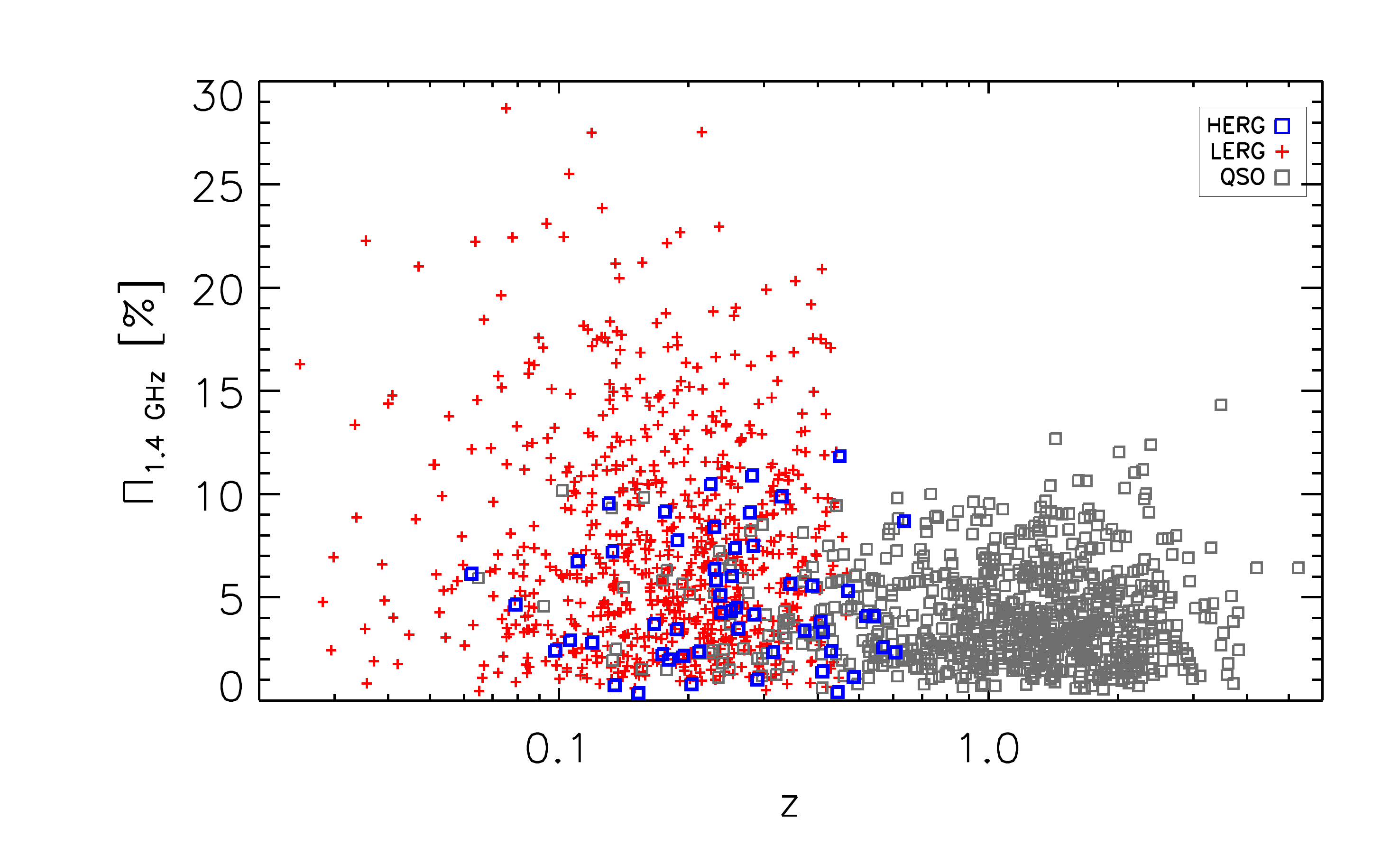}
      \caption{Integrated degree of polarization at 1.4~GHz ($\Pi_{\rm 1.4\,GHz}$) versus 
      redshift ($z$) for LERGs (red), HERGs (blue, hatched) and radio-loud QSOs (grey).  
}
  \label{fpol_z}
\end{figure}

\subsection{Radio Luminosity Distribution of Polarized HERGs and LERGs}

\begin{figure}
\centering
\vspace{-0.5cm}
    \includegraphics[width=8.5cm]{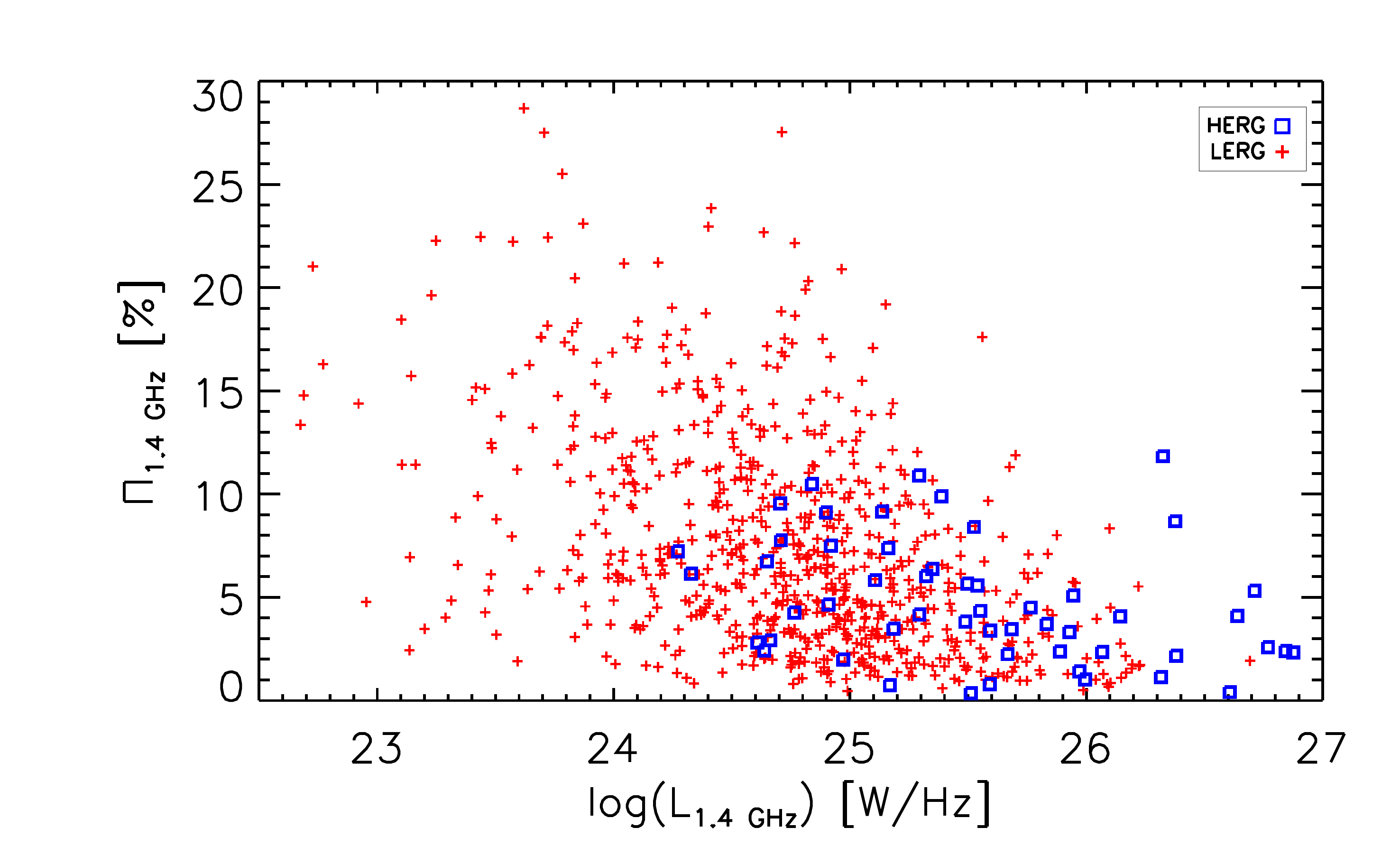}
      \caption{Integrated degree of polarization at 1.4~GHz ($\Pi_{\rm 1.4\,GHz}$) versus
      the 1.4~GHz radio luminosity, in units of W~Hz$^{-1}$, for HERGs (blue squares)
      and LERGs (red plus symbols). 
      Note that we are insensitive to many low-luminosity, low-$\Pi_{\rm 1.4\,GHz}$ 
      sources due to the polarization detection limit. 
}
  \label{lum_herg_lerg}
\end{figure}

The distribution of $\Pi_{\rm 1.4\,GHz}$ with 1.4 GHz radio luminosity ($L_{\rm 1.4\,GHz}$) 
is shown in Figure~\ref{lum_herg_lerg}. We calculate the radio luminosity using 
$L_{\rm 1.4\,GHz}=4\pi D_L^2 S_{\rm 1.4 GHz} (1+z)^{-(\alpha+1)}$,
where $D_L$ is the luminosity distance, $S_{\rm 1.4 GHz}$ is the integrated total intensity 
and we have assumed a spectral index $\alpha=-0.7$.  
\footnote{We define the spectral index, $\alpha$, 
such that the observed flux density ($S$) at frequency $\nu$ follows the relation 
$S_{\nu}\propto\nu^{\rm{+}\alpha}$.}
In the local universe, LERGs dominate the radio-loud AGN number counts at 
low luminosities ($22\lesssim \log(L_{\rm 1.4\,GHz}~{\rm [W~Hz^{-1}]}) \lesssim 25$), while 
above $L_{\rm 1.4\,GHz}\sim10^{26}~{\rm W~Hz^{-1}}$, HERGs begin to dominate (BH12). 
While examples of both populations are found across the full range of HERG and LERG luminosities, 
we only detect polarized HERGs above $L_{\rm 1.4\,GHz}\sim10^{24}~{\rm W~Hz^{-1}}$. 

Since LERGs are more numerous than HERGs at lower luminosity and we have just 
shown that LERGs can achieve higher values of $\Pi_{\rm 1.4\,GHz}$, then one might 
expect an anti-correlation between $\Pi_{\rm 1.4\,GHz}$ and $L_{\rm 1.4\,GHz}$. 
We are insensitive to low-luminosity low-$\Pi_{\rm 1.4\,GHz}$ sources 
due to the polarization detection limit, and this strongly affects any robust inferences 
on any dependence of $\Pi_{\rm 1.4\,GHz}$ on $L_{\rm 1.4\,GHz}$. 
If we only consider sources with total intensities greater than 50~mJy, we have 
roughly 70\% of both HERG and LERG sources below the detection limit in 
polarized flux density. This is roughly the limit at which useful statistics can be derived 
from data with upper limits (e.g.~Antweiler \& Taylor 2008). 
Using a Kendall's tau rank correlation, we find a statistically significant (p-value of $6\times10^{-8}$, $\sim$$5.4\sigma$),
but weak, anti-correlation (correlation coefficient of $-0.1$) between $\Pi_{\rm 1.4\,GHz}$ and $L_{\rm 1.4\,GHz}$ 
for all sources. 
Considering the HERG sources only, we find no statistically significant correlation 
but we find a marginally significant anti-correlation between $\Pi_{\rm 1.4\,GHz}$ and 
$L_{\rm 1.4\,GHz}$ for the LERG sources 
(p-value of 0.013\%, $\sim$$3.8\sigma$, for a correlation coefficient of $-0.1$).  
There is no evidence for an anti-correlation between $\Pi_{\rm 1.4\,GHz}$ and $L_{\rm 1.4\,GHz}$ 
for the brightest sources (i.e.~with Stokes $I>100$~mJy, as shown in Appendix A). 
This suggests that any anti-correlation between $\Pi_{\rm 1.4\,GHz}$ and $L_{\rm 1.4\,GHz}$ 
is mainly driven by the change in the dominant population of radio-loud AGN 
towards lower luminosities from HERGs to LERGs. 
Much deeper radio polarization observations are required to more robustly 
test the existence of an anti-correlation between $\Pi_{\rm 1.4\,GHz}$ and 
$L_{\rm 1.4\,GHz}$ for the LERG sources. Of course, radio luminosity may 
be anti-correlated with $\Pi_{\rm 1.4\,GHz}$ independent of the HERG/LERG classification 
and deeper polarization observations would also provide a key test for the continued absence 
of low-luminosity HERGs with $\Pi_{\rm 1.4\,GHz}>15\%$.

\begin{figure}
\centering
\vspace{-0.5cm}
    \includegraphics[width=8.5cm]{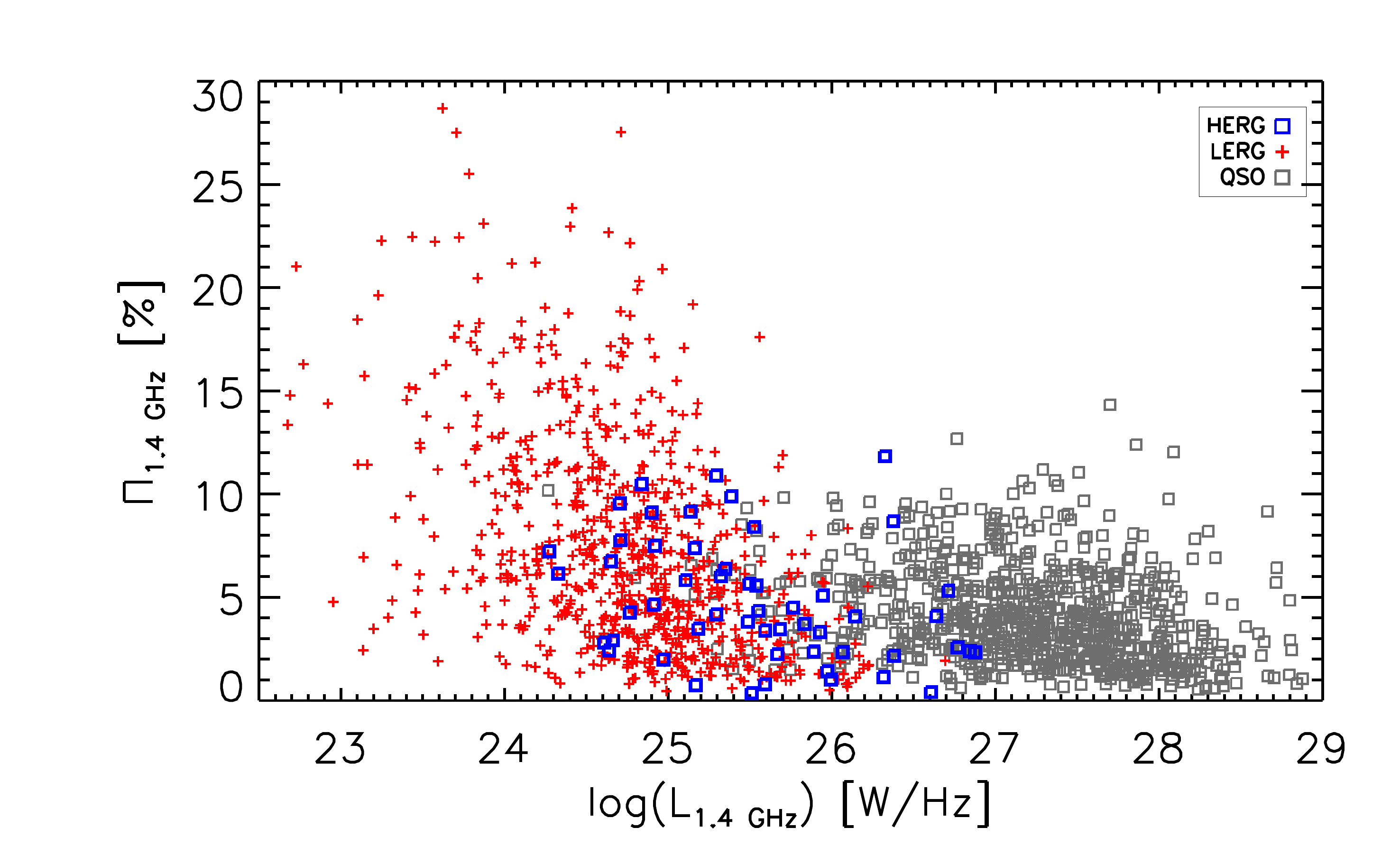}
      \caption{Integrated degree of polarization at 1.4~GHz ($\Pi_{\rm 1.4\,GHz}$) versus
      the 1.4~GHz radio luminosity, in units of W~Hz$^{-1}$, for HERGs (blue squares)
      and LERGs (red plus symbols) and radio-loud QSOs (grey squares). 
}
  \label{lum_herg_lerg_qso}
\end{figure}

It is interesting to extend this plot to higher radio luminosities by including all 
the known radio polarized QSOs (Hammond et al.~2012) that overlap with the sky coverage of our 
sample (i.e.~the SDSS area). Figure~\ref{lum_herg_lerg_qso} shows $\Pi_{\rm 1.4\,GHz}$ versus 
$L_{\rm 1.4\,GHz}$ again but with the QSOs included (grey squares). This 
demonstrates the extension of the radiative-mode AGN population to higher 
luminosities and clearly shows the limitation of radiative-mode AGN to 
values of  $\Pi_{\rm 1.4\,GHz}<15\%$, across five orders of magnitude in radio luminosity.  

\begin{figure}
\centering
\vspace{-0.5cm}
    \includegraphics[width=8.5cm]{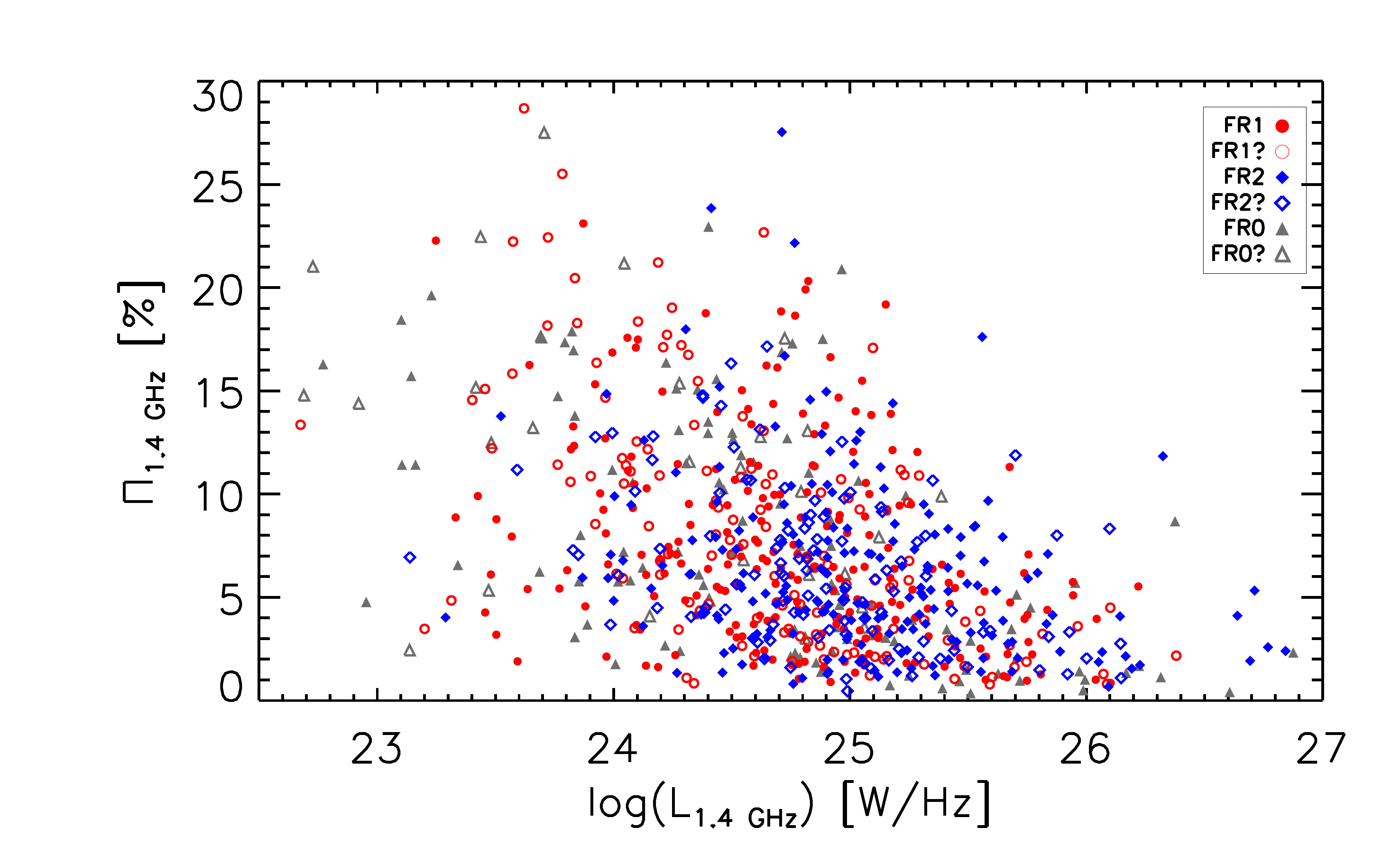}
      \caption{Integrated degree of polarization at 1.4~GHz ($\Pi_{\rm 1.4\,GHz}$) versus
      the 1.4~GHz radio luminosity, in units of W~Hz$^{-1}$, for FR0 (grey triangles), 
      FR1 (red circles) and FR2 (blue diamonds) radio morphologies. Filled symbols 
      represent robust classifications, with open symbols representing uncertain classifications. 
      See Sections 2.3 and 3.3.1 for details.  
}
  \label{lum_morph}
\end{figure}

\begin{figure}
\centering
\vspace{-0.5cm}
    \includegraphics[width=8.5cm]{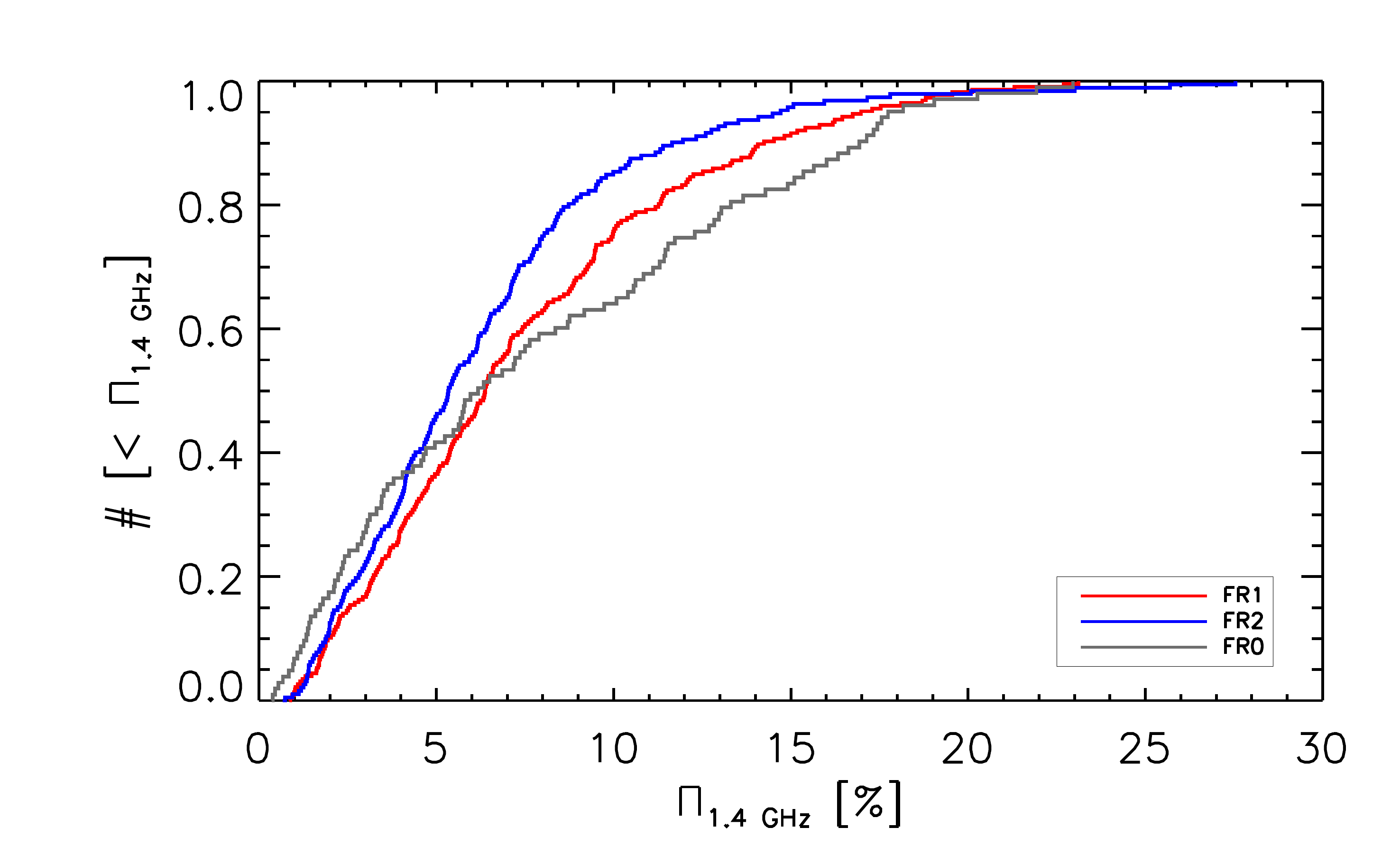}
      \caption{Empirical cumulative distribution functions (ECDFs) of the integrated degree of 
      polarization ($\Pi_{\rm 1.4\,GHz}$), for FR2 (blue), FR1 (red) and FR0 (grey) 
      classifications. 
}
  \label{ECDF_fr}
\end{figure}

\begin{figure}
\centering
\vspace{-0.5cm}
    \includegraphics[width=8.5cm]{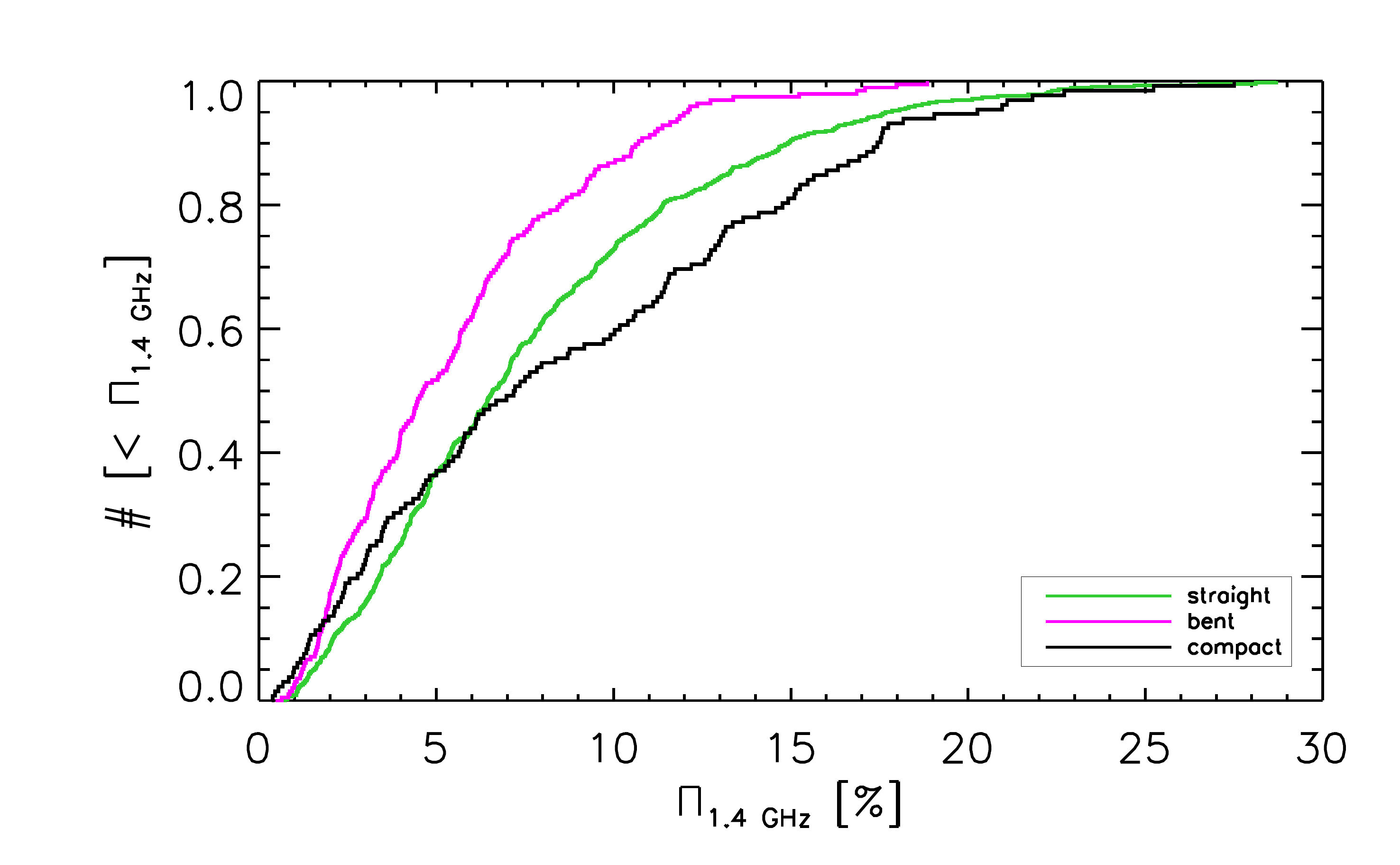}
      \caption{Empirical cumulative distribution functions (ECDFs)
      of the integrated degree of polarization at 1.4~GHz ($\Pi_{\rm 1.4\,GHz}$) 
      for radio morphologies classified as straight (green), bent (magenta) and compact (black).  
      See Sections 2.3 and 3.3.2 for details.  
}
  \label{hist_sbc}
\end{figure}

\begin{figure}
\centering
    \includegraphics[width=8.5cm]{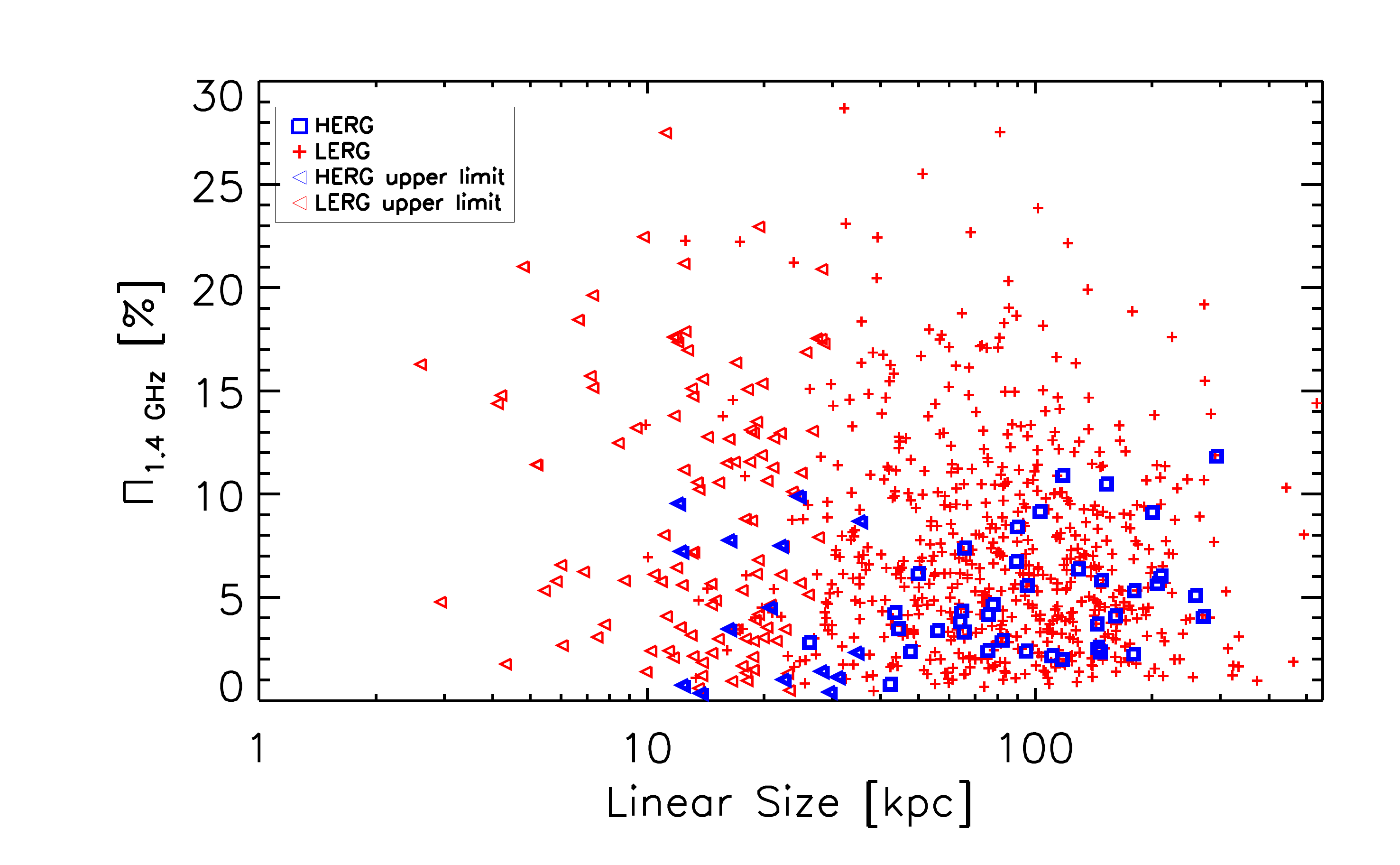}
      \caption{Integrated degree of polarization at 1.4~GHz ($\Pi_{\rm 1.4\,GHz}$) versus
      projected linear source size in kpc, for LERGs (red plus symbols, with upper limits 
      denoted by red triangles) and HERGs (blue squares, with upper limits denoted by 
      blue triangles). See Sections 2.3 and 3.3.3 for details. 
}
  \label{linear_size}
\end{figure}

\begin{figure}
\centering
    \includegraphics[width=8.5cm]{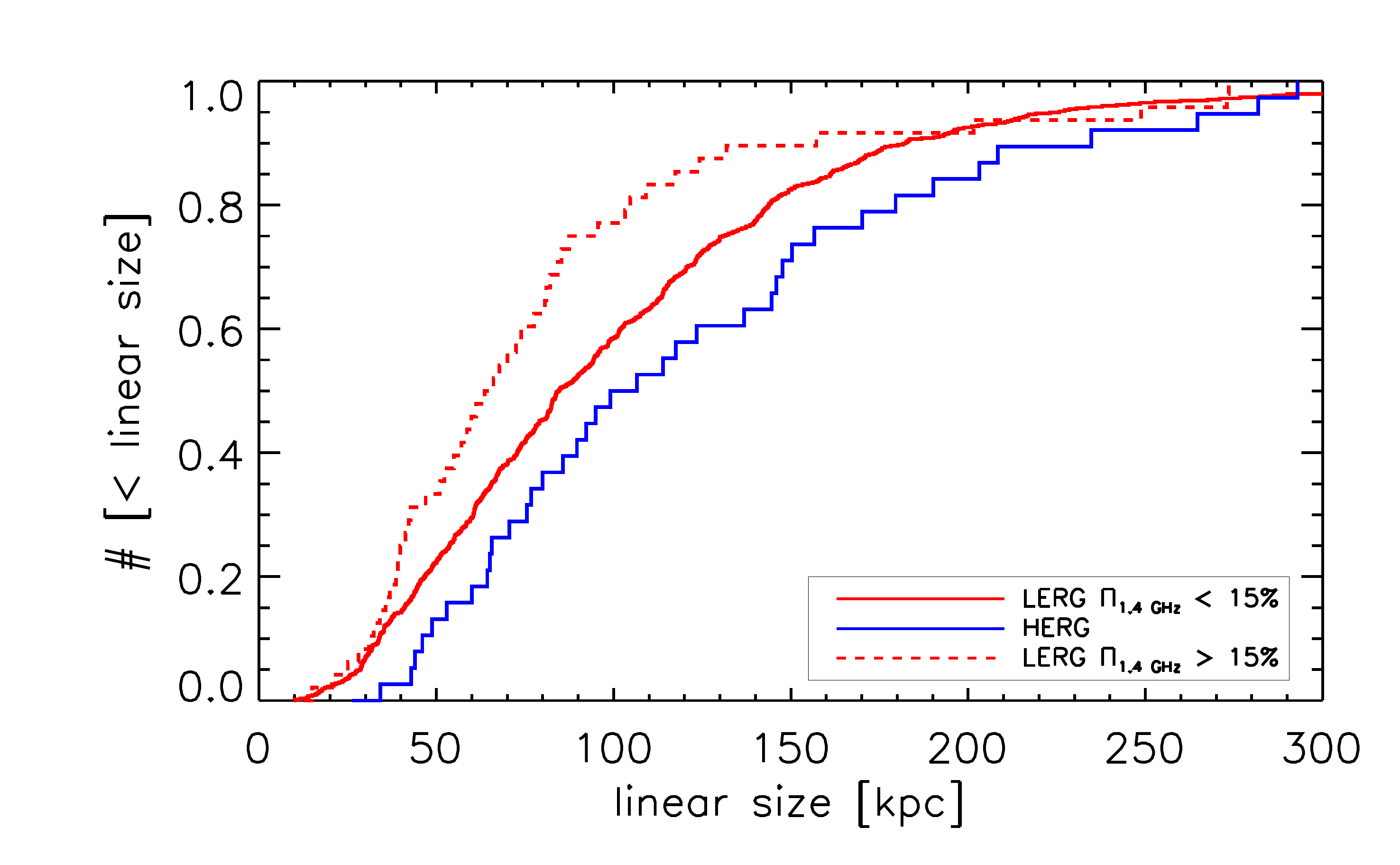}
      \caption{Empirical cumulative distribution function (ECDF) of the source linear size in 
      kpc (excluding sources with upper limits), for LERGs with $\Pi_{\rm 1.4\,GHz}<15\%$ 
      (red solid line), HERGs (blue solid line) and LERGs with $\Pi_{\rm 1.4\,GHz}>15\%$ (red dashed line). 
}
  \label{ECDF_linear_size}
\end{figure}

\subsection{Radio morphology and integrated degree of polarization}

We now present an investigation of the origin of the difference in 
$\Pi_{\rm 1.4\,GHz}$ for HERGs and LERGs in relation to the morphological 
properties of the large scale radio emission (total intensity only, from the FIRST survey).  

\subsubsection{FR0, FR1 and FR2 }
Using our manual classification of FR0/FR1/FR2, described in Section 2.3, we find 14/6/18 HERGs that 
are FR0/FR1/FR2 with 15 uncertain classifications and 89/221/174 LERGs that are FR0/FR1/FR2 with 255 having 
an uncertain classification. It is immediately clear from this that the HERG/LERG class is essentially 
independent of the FR type. This is consistent with other studies by Best et al.~(2009), 
Lin et al.~(2010) and Gendre et al.~(2013).  
Figure~\ref{lum_morph} again plots $\Pi_{\rm 1.4\,GHz}$ versus radio luminosity but now with the FR-type 
identified. It is clear from this plot that the range of $\Pi_{\rm 1.4\,GHz}$ is not restricted to values 
less than $15\%$ for FR1, FR2 or FR0 sources. 
Thus, the HERG-LERG difference in $\Pi_{\rm 1.4\,GHz}$ cannot be easily explained by the radio-morphology type.  
Similar results are obtained if we only consider the brightest sources, with $I>100$~mJy (see Appendix A).
Figure~\ref{ECDF_fr} shows the empirical 
cumulative distribution functions (ECDFs) for $\Pi_{\rm 1.4\,GHz}$ of FR0, FR1 and FR2 sources. 
Considering only the definitive FR1 and FR2 type sources, we find a marginally significant difference ($\sim$2.3$\sigma$)
in $\Pi_{\rm 1.4\,GHz}$ between the two types, with FR1s having a median $\Pi_{\rm 1.4 GHz, FR1}$ of 6.4\% 
and a median $\Pi_{\rm 1.4 GHz, FR2}$ of 5.3\% for FR2s. Similar results were previously found by Grant (2011) for 
a smaller number of sources (19 FR1s and 17 FR2s). We find no significant difference in 
$\Pi_{\rm 1.4\,GHz}$ between FR0 and FR1 sources (p-value of 18\%) but get a $\sim$$3\sigma$ difference 
(p-value of 0.3\%) in $\Pi_{\rm 1.4\,GHz}$ between FR0 and FR2 sources. 
The median $\Pi_{\rm 1.4 GHz, FR0}$ of FR0 sources is 6.1\%.  

\subsubsection{Straight jets and bent jets}
Intrinsic polarization angle cancellation in radio sources with bent or disturbed morphologies 
could result in low integrated degrees of polarization, while the undisturbed, straight 
jet sources may achieve much higher values of $\Pi_{\rm 1.4\,GHz}$. 
Figure~\ref{hist_sbc} presents ECDFs 
of $\Pi_{\rm 1.4\,GHz}$ split into straight, bent and compact sources 
(as defined in Section 2.3). The majority of polarized sources are classified as straight (58\%), with 
25\% bent and the remaining 17\% being compact. For sources with $\Pi_{\rm 1.4\,GHz} > 15\%$ 
the fraction of straight sources is 60\% (similar to the fraction for all sources), however, the fraction 
of bent sources falls to only 5\%. KS-tests show that both compact and straight sources have 
significantly different $\Pi_{\rm 1.4\,GHz}$ distributions than bent sources ($\sim$$4.7\sigma$ and 
$4.3\sigma$, respectively) while compact and straight sources do not differ significantly from each 
other (p-value of 3\%). We find that 40\% of HERGs are straight, 32\% are bent 
and 28\% are compact, while 60\% of LERGs are straight with 24\% bent and 16\% compact. 

\subsubsection{Influence of the `radio core'}
The contribution of radio emission from active or `restarted' emission regions near the 
central engine (i.e.~the radio core) could decrease the integrated degree of polarization. 
This is possible because the inner regions of the radio source should suffer larger 
amounts of depolarization due to being more deeply embedded in the host galaxy. 
Furthermore, synchrotron self-absorbed and optically thick emission regions can only 
reach intrinsic degrees of polarization of approximately 10\% (e.g.~Pacholczyk 1970). 
In this case the additional total flux density provided by the core is likely to contribute 
very little additional polarized flux, causing the overall degree of polarization to decrease. 
To investigate whether or not this was an important effect for our sample, we determined 
the number of extended sources (i.e.~with angular sizes $>$10") that also had radio cores 
in the FIRST images. 
Out of the 38 polarized HERGs with angular sizes greater than 10", 6 of them have radio 
cores ($\sim$16\%). The median value of $\Pi_{\rm 1.4\,GHz}$ for these 6 sources is $3.7\%$, 
while the median value of $\Pi_{\rm 1.4\,GHz}$ for the other 32 extended sources without a core is 4.4\%.
This indicates that the presence of a core may have a small effect in reducing $\Pi_{\rm 1.4\,GHz}$ 
for HERGs (although a two-sided KS test finds no significant difference 
in $\Pi_{\rm 1.4\,GHz}$ for HERGs with and without a core, p-value of 58\%). 
Of the 597 extended LERGs, 137 have radio cores ($\sim$23\%). This shows that the presence 
of a core is slightly more likely for LERGs that HERGs. The median value of $\Pi_{\rm 1.4\,GHz}$ 
for LERGs with cores is 6.6\% and without cores is slightly lower, at 6.1\%. 
In fact, there are 17 LERGs with radio cores that have $\Pi_{\rm 1.4\,GHz}>15\%$, 
showing that the presence of a core is not a hindrance to high integrated degrees of polarization; 
these cores may be dominated by bright, highly polarized, optically-thin inner jet regions.

\subsubsection{Projected linear size}
One might expect the extended polarized emission of sources with larger 
linear sizes to have less Faraday depolarization, and thus higher $\Pi_{\rm 1.4\,GHz}$, 
due to the propagation of the radiation through a presumably less dense magnetoionic 
environment (e.g.~Hardcastle \& Krause 2014). 
To investigate this possibility, we compare the projected linear extent of the radio emission 
in the source rest frame, measured as described in Section 2.3, to $\Pi_{\rm 1.4\,GHz}$,
for the HERG and LERG sources (Figure~\ref{linear_size}). 
For all sources, there is no obvious strong dependence of $\Pi_{\rm 1.4\,GHz}$ on linear size. 
Excluding the sources with upper limits, the median linear size of the HERG sample is 102.8~kpc, 
and 82.8~kpc for the LERG sample. Thus, if the Hardcastle \& Krause (2014) simulation results 
were true for all sources, then one would expect $\Pi_{\rm 1.4\,GHz}$ for HERGs to extend 
to higher values than the LERGs, completely the opposite to what is observed. 
A two-sided KS test indicates, however, that the linear size distribution of the HERGs is not 
significantly different from that of the LERGs (p-value of 17\%). 
The median linear size of sources with $\Pi_{\rm 1.4\,GHz}>15\%$ is 64.7~kpc, with 
a KS-test indicating a significant difference in linear size compared to sources with 
$\Pi_{\rm 1.4\,GHz}<15\%$ (p-value of 0.6\%, $\sim$2.7$\sigma$). 
This surprisingly indicates that the most highly polarized LERGs are the ones with smaller 
projected linear extents (Figure~\ref{ECDF_linear_size}). 
If we include the upper limits as measurements then the median linear size for HERGs decreases 
to 75.4~kpc and 72.0~kpc for LERGs. The larger decrease for HERGs occurs because a greater 
fraction of HERG sources are compact (c.f.~Section 3.3.2). 

We caution that the smaller linear size estimates in the fainter NVSS sources may be 
confounded by a smaller estimate of the angular size due to the extended emission being 
resolved out/undetected in FIRST. 
Furthermore, while the measurement of $\Pi_{\rm 1.4\,GHz}$ is independent of the FIRST 
angular size measurement, it is worth considering here whether or not the NVSS 
angular size influences the measurement of $\Pi_{\rm 1.4\,GHz}$ differently for HERGs or 
LERGs, even though the majority of sources are unresolved (84\%). 
However, a KS test of NVSS angular size for HERGs and LERGs shows no significant difference (p-value of 26\%).

\subsection{Environmental diagnostics of HERGs and LERGs}

Differences in the large scale gaseous and magnetoionic environments are prime candidates 
for explaining the difference in $\Pi_{\rm 1.4\,GHz}$ for HERGs and LERGs, since the radio jet
emission and morphology is expected to be strongly influenced by its local environment (e.g.~Laing et al.~2014). 
Here we explore several potential tracers of the environment local to the source. 

\subsubsection{Faraday Rotation Measure}
The integrated fractional polarization at 1.4~GHz is expected to be strongly affected 
by Faraday rotation, where magnetoionic material along the line of sight causes a 
rotation of the intrinsic linear polarization angle ($\Psi$) with wavelength squared, 
$\Delta\Psi = {\rm RM}\,\lambda^2$. 
The Faraday rotation measure (RM) is defined as $0.81 \int^{0}_{L}{ n_e {B_{||}} dl} ~~{\rm rad~m}^{-2}$, 
where $n_e$ is the electron number density in cm$^{-3}$, $B_{||}$ is the line-of-sight 
magnetic field strength in $\mu$G and $L$ is the distance through the magnetoionic region in parsecs. 
For example, with $B_{||}\sim1$~$\mu$G, $n_e\sim10^{-4}$~cm$^{-3}$ and $L\sim100$~kpc 
gives an RM of $\sim$$\pm8$~rad~m$^{-2}$, which causes a polarization angle 
rotation of 20 degrees at 1.4~GHz. Fluctuations in RM of this order (and larger) across the jets and 
lobes of radio galaxies are not uncommon (e.g.~Laing et al.~2008, Guidetti et al.~2012). 
Thus, we expect such RM fluctuations local to the radio source 
to strongly reduce the integrated fractional polarization at 1.4 GHz. 

\begin{figure}
\centering
\vspace{-0.5cm}
    \includegraphics[width=8.5cm]{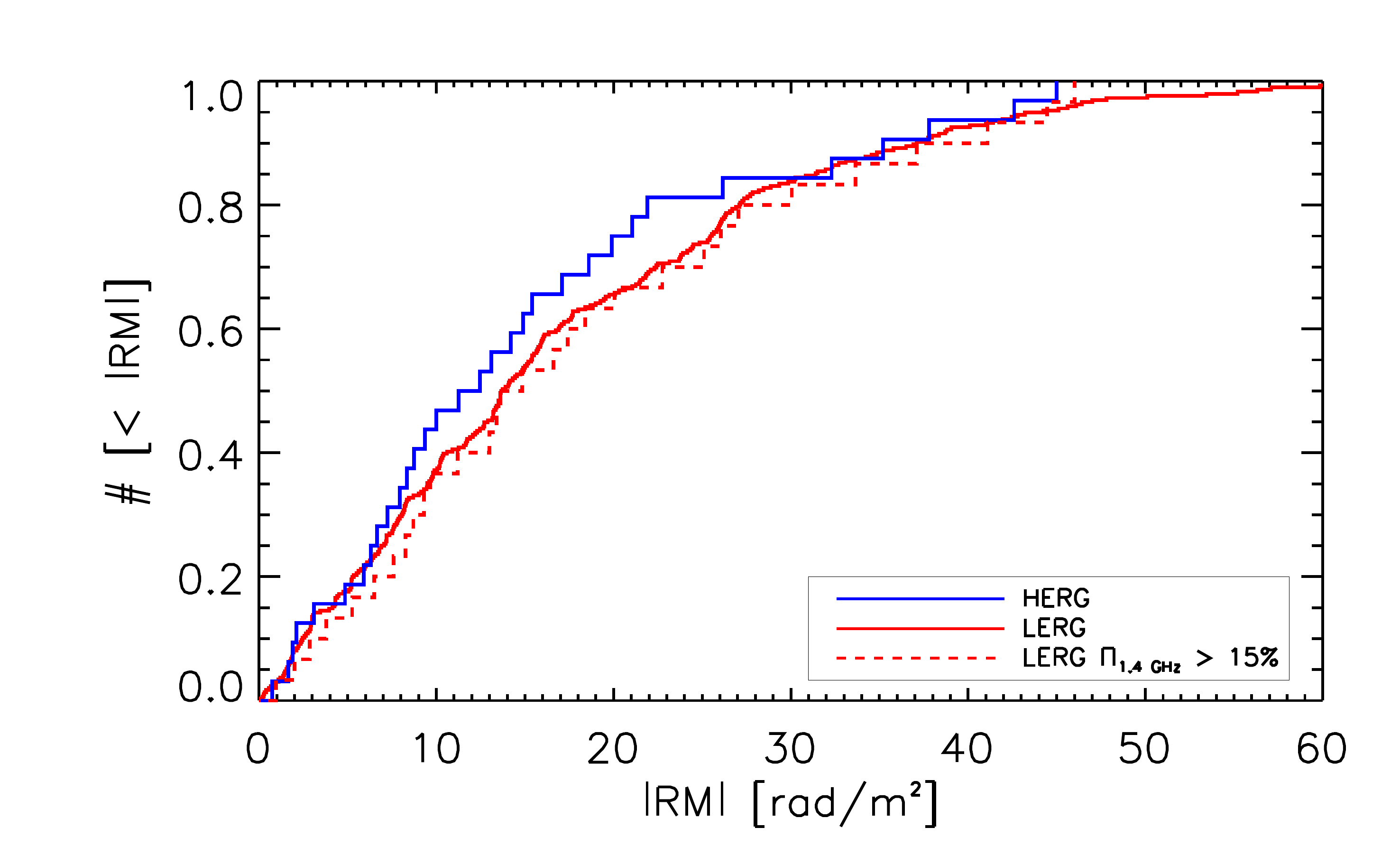}
      \caption{Empirical cumulative distribution function (ECDF) of the absolute value of the 
      Faraday rotation measure ($|$RM$|$) in rad/m$^{-2}$, for LERGs (red solid line), HERGs (blue solid line) 
      and LERGs with $\Pi_{\rm 1.4\,GHz}>15\%$ (red dashed line). 
}
  \label{rm}
\end{figure}

In order to investigate the relative importance of this for HERGs and LERGs, we use 
the RM catalog of Taylor et al.~(2009), that produced 37,543 RMs by reprocessing the NVSS data into two 
channels centred on 1.365 and 1.435 GHz. Using this catalog, we can compare the RM 
of the integrated polarized emission for our sample of HERGs and LERGs. 
In Figure~\ref{rm}, we plot the empirical cumulative distribution functions (ECDFs) of the absolute value of 
the RMs for 295 LERGs and 34 HERGs\footnote{The reason we did not find RMs for 
all our sources is that the Taylor et al.~(2009) RM catalog uses a higher threshold of $8\sigma_{QU}$ ($\sim$2.3~mJy).}. 
A two-sided KS test finds no significant difference between the two distributions (p-value of 95\%). 
Furthermore, we also find no significant difference between the $|$RM$|$ of sources 
with $\Pi_{\rm 1.4\,GHz}>15\%$ and those with $\Pi_{\rm 1.4\,GHz}<15\%$ (p-value of 99\%). 
We searched for differences in $|$RM$|$ between FR types, between straight and bent 
sources and in relation to linear size, finding no significant difference or dependence on 
any of the sub-samples. 
This is somewhat unsurprising since we know that the RMs from Taylor et al.~(2009) are 
dominated by the Galactic contribution (Schnitzeler et al.~2010, Stil et al.~2011). Attempts to 
model the Galactic RM component and subtract it from the total RM to robustly recover the extragalactic 
contribution is not possible with the current limited data due to poor constraints on individual 
source errors (e.g.~Oppermann et al.~2015). However, this is expected to be possible in future 
with higher precision RM measurements from observations with much broader and 
more continuous wavelength-squared coverage.

\subsubsection{Depolarization due to the local environment}
As described above, fluctuations in the magnetoionic medium across the source 
can significantly depolarise the radio emission. 
The recent study of Farnes et al.~(2014) presented a parametrization of the 
change in the degree of polarization with wavelength (i.e.~a degree of polarization 
spectral index, $\beta$, where $\Pi_\lambda \propto \lambda^\beta$) for 951 radio sources. They found that sources
with a steep total intensity spectrum exhibited depolarization (a decreasing degree of 
polarization with increasing wavelength), while flat-spectrum sources typically maintain 
approximately constant degrees of polarization over large ranges in wavelength. 
From this they concluded that $\beta$ is predominantly affected by the magnetoionic 
environment local to the source, rather than by material in the distant foreground (i.e.~intergalactic 
magnetic fields and/or the Milky Way). Only twelve of the sources presented here 
have an estimate of $\beta$ from Farnes et al.~(2014), with $\beta$ ranging from $-0.8$ to 
$0.3$, with a median value of $-0.2$. While this is insufficient for a robust 
statistical analysis of our sample, we can use the overall results of Farnes et al.~(2014) 
to infer that $\Pi_{\rm 1.4\,GHz}$ for the majority of our sources, which have steep total intensity 
spectral indices, are likely affected by depolarization due to the local environment. 
Further observations are required to determine if HERGs and LERGs have 
different depolarisation properties.

\subsubsection{Galaxy number density}
We now consider the polarization properties of HERGs and LERGs in comparison with 
one of the most direct measure of environment, galaxy number density. 
Estimates of the galaxy number density are available for $\sim$20\% of our 
polarized sources, from Lin et al.~(2010). They estimated the excess number of galaxies 
over the mean background within 0.5~Mpc ($\Sigma_{\rm 0.5\,Mpc}$) of the host galaxies 
of 1040 extended FIRST sources. 
For the polarized HERGs and LERGs, the median value of $\Sigma_{\rm 0.5\,Mpc}$ is 6.6, 
compared to the unpolarized sources, that have a median $\Sigma_{\rm 0.5\,Mpc}$ of $8.8$, with 
a KS test indicating a significant difference (p-value of $4.8\times10^{-5}$, $\sim$4$\sigma$).
This means that the polarized sources are typically in more underdense environments, in comparison
to the unpolarised sources. 
The median value of $\Sigma_{\rm 0.5\,Mpc}$ for all (polarized and unpolarized) HERGs 
is 4.1 compared to 8.6 for all LERGs, supporting previous studies that found that HERGs 
are generally in less dense environments that LERGs (e.g.~Best 2004). 
A KS test of $\Sigma_{\rm 0.5\,Mpc}$ for polarized HERGs versus polarized LERGs find no 
significant difference between the sources (p-value of 10\%). However, only five polarized HERGs have 
$\Sigma_{\rm 0.5\,Mpc}$ estimated, with all having values of $\Sigma_{\rm 0.5\,Mpc}<6.8$. 
For the LERGs, the median values of $\Pi_{\rm 1.4\,GHz}$ are 6.3\% and 7.5\% for sources with 
$\Sigma_{\rm 0.5\,Mpc}$ greater than and less than the median value of $\Sigma_{\rm 0.5\,Mpc}$, 
respectively (p-value of 0.3\%, $\sim$2.9$\sigma$ significance). This suggests that, at least for the LERGs, 
the highest integrated degrees of polarization appear to favour the lowest galaxy density environments.

\subsubsection{Spectral index as an environment density probe}

Samples of radio galaxies selected by spectral index have previously shown that the 
steepest spectral index sources typically reside in rich clusters of galaxies 
(Baldwin \& Scott 1973; Slee, Siegman \& Wilson 1983). This has been interpreted as a 
result of pressure-confinement of the radio lobes propagating in dense environments (e.g.~Klamer et al.~2006). 
We have total intensity spectral index measurements ($\alpha$) from Farnes et al.~(2014) 
for 32\% of LERGs and 56\% of HERGs, with 95\% of LERGs and 100\% of 
HERGs having $\alpha$ within the range $-1.5 < \alpha < -0.3$. This supports the 
expectation that the majority of our polarized sources are indeed lobed radio galaxies 
with the dominant polarized emission presumably coming from the steep spectrum, extended 
regions of the source (c.f.~Banfield et al.~2011, Hales et al.~2014).  

\begin{figure}
\centering
\vspace{-0.5cm}
    \includegraphics[width=8.5cm]{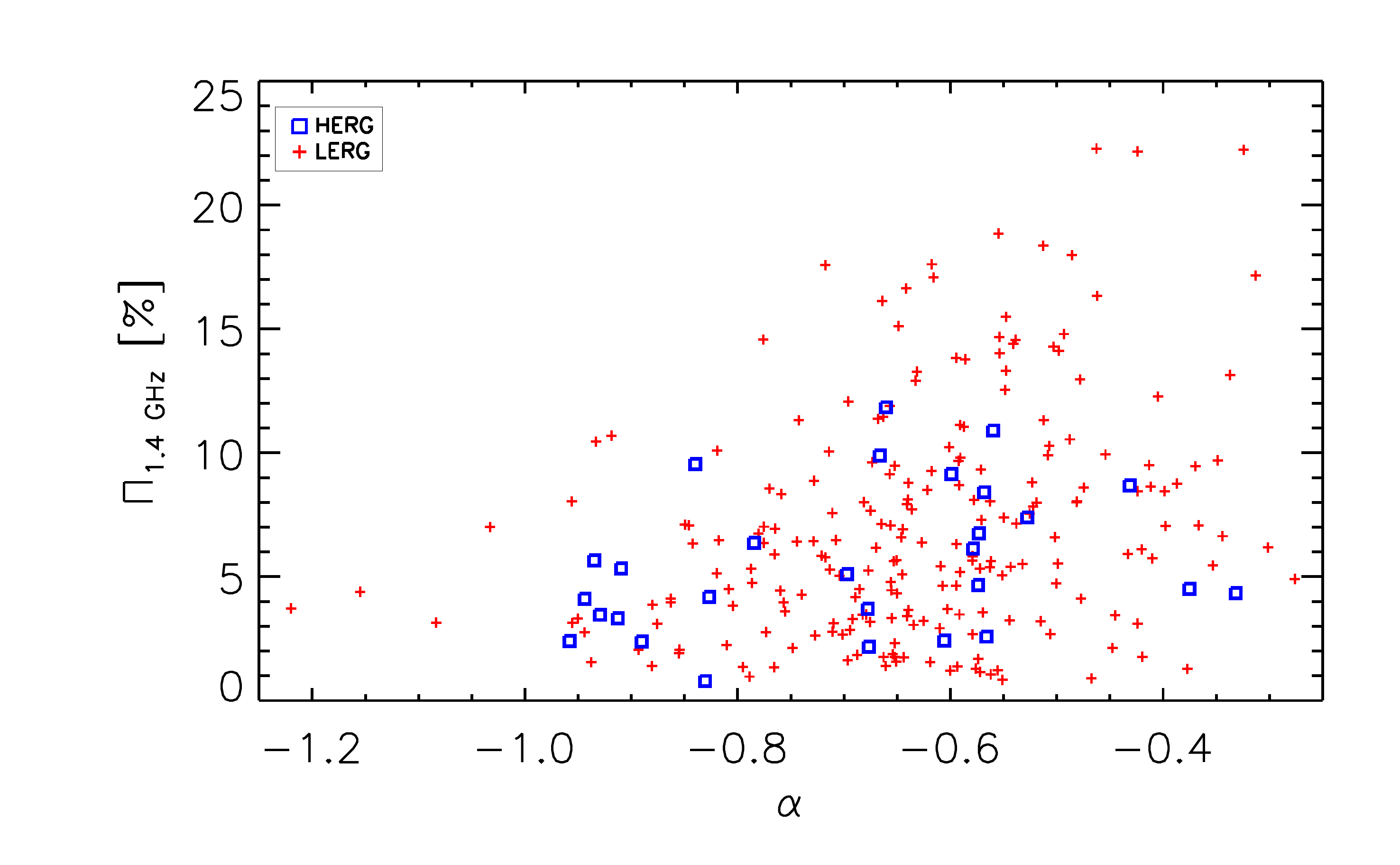}
      \caption{Integrated degree of polarization at 1.4~GHz ($\Pi_{\rm 1.4\,GHZ}$) 
      versus the total intensity spectral index ($\alpha$) taken from Farnes et al.~(2014), 
      for HERGs (blue squares) and LERGs (red plus symbols). See Section~4.1.2 for details.
}
  \label{spix}
\end{figure}

From the plot of $\Pi_{\rm 1.4\,GHz}$ versus $\alpha$, shown in Figure~\ref{spix}, one can 
immediately notice the absence of sources with large $\Pi_{\rm 1.4\,GHz}$ and steep 
$\alpha$ (i.e.~$\alpha<-0.7$). However, this appears to be only true for the LERG sources 
since the HERGs essentially uniformly cover their $\Pi_{\rm 1.4\,GHz}$--$\alpha$ parameter space. 
Using a Kendall's tau rank correlation test, we find a significant correlation between $\alpha$ and 
$\Pi_{\rm 1.4\,GHz}$ for the steep-spectrum LERGs, with a correlation coefficient of 0.2 at $\sim$4.5$\sigma$ significance. 
Employing the same correlation test for the HERGs only, finds no significant correlation. 
There are no known selection effects in Farnes et al.~(2014) that could connect both the spectral 
index (derived from multiple surveys), and the polarisation fraction (derived from just NVSS).
Therefore, if very steep values of $\alpha$ are indicative of dense environments, 
then it is interesting to relate the absence of highly polarized LERG sources with 
$\alpha<-0.7$ to the effect of the environment.

\subsection{Host galaxy and accretion rate properties of polarized HERGs and LERGs}

In order to infer general properties of the HERG and LERG populations from the subset 
of sources with significant integrated polarized emission at 1.4 GHz, we need to 
assess whether or not the polarized sources are generally representative of their 
individual classes. Previous studies have found that in the local universe HERGs 
are generally found in disk galaxies with ongoing star formation and with a high density of cold 
gas accreting at a rate greater than $\sim$1\% of the Eddington rate. 
In contrast, LERGs are mainly found in massive elliptical galaxies with more 
massive black holes fuelled by slowly cooling hot gas that accretes at less than 
1\% of Eddington. The accretion energy of LERGs is mainly channeled into 
their jets, that in turn limit the amount of hot gas cooling in their host galaxy 
and cluster environments in a feedback cycle. See Heckman \& Best (2014) for a 
comprehensive review. In the remaining part of this section, we present the results of
our analysis of the optical, infrared and radio properties of the polarized HERG/LERG 
sample in order to determine their host galaxy and accretion rate properties.

\subsubsection{WISE Colour-Colour Host Galaxy Diagnostics}
\begin{figure}
\centering
\vspace{0.5cm}
    \includegraphics[width=8.5cm]{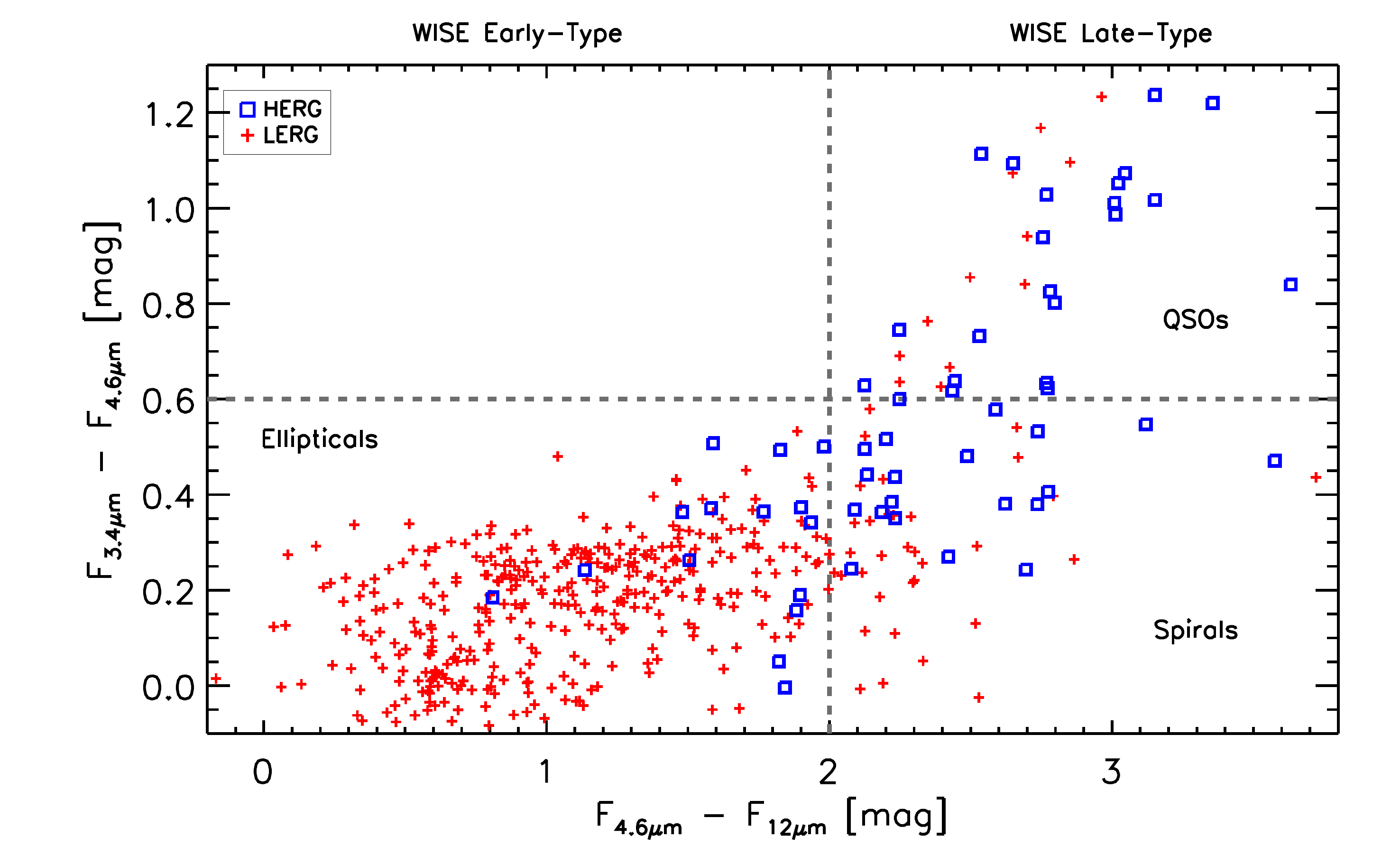}
      \caption{WISE infrared colour-colour plot for LERGs (red plus symbols) 
      and HERGs (blue squares). The dashed lines denote the regions 
      where different galaxy populations are expected to be. See Section 3.5.1 
      for more details. 
}
  \label{wise}
\end{figure}

The IR properties across the WISE band can be used to help separate early-type 
elliptical galaxies from galaxies with a substantial amount of on-going star formation 
in disky/spiral galaxies as well as identifying those AGN with a bright accretion disk (i.e.~a QSO). 
Figure~\ref{wise} shows the WISE colour-colour plot for our sample. 
The dotted horizontal and vertical lines are from Wright et al. (2010), who divide elliptical 
and spiral galaxies at a WISE [4.6]--[12] colour of $+1.5$~mag and find that the most powerful 
optical AGN lie above a [3.4]--[4.6] colour of $+0.6$~mag. 
From this we can see that the majority of LERGs are hosted by elliptical galaxies with only 
a small fraction classified as having spiral-type hosts. The LERGs classified as QSOs are 
potentially heavily obscured Type~2 AGN. 
The polarized HERGs are mainly in the WISE late-type category (72\%) with most of the rest 
(22\%) clustered within 0.5 mag of the dividing line between WISE early and late-type 
categories. 
This is broadly consistent with the BH12 detailed study of the optical host galaxy properties, 
where HERG host galaxies were typically of lower stellar mass, lower black 
hole mass (i.e.~less evolved) and had bluer colours than the LERGs. 

\subsubsection{WISE Mid-IR Luminosity}
\begin{figure}
\centering
\vspace{-0.5cm}
    \includegraphics[width=8.5cm]{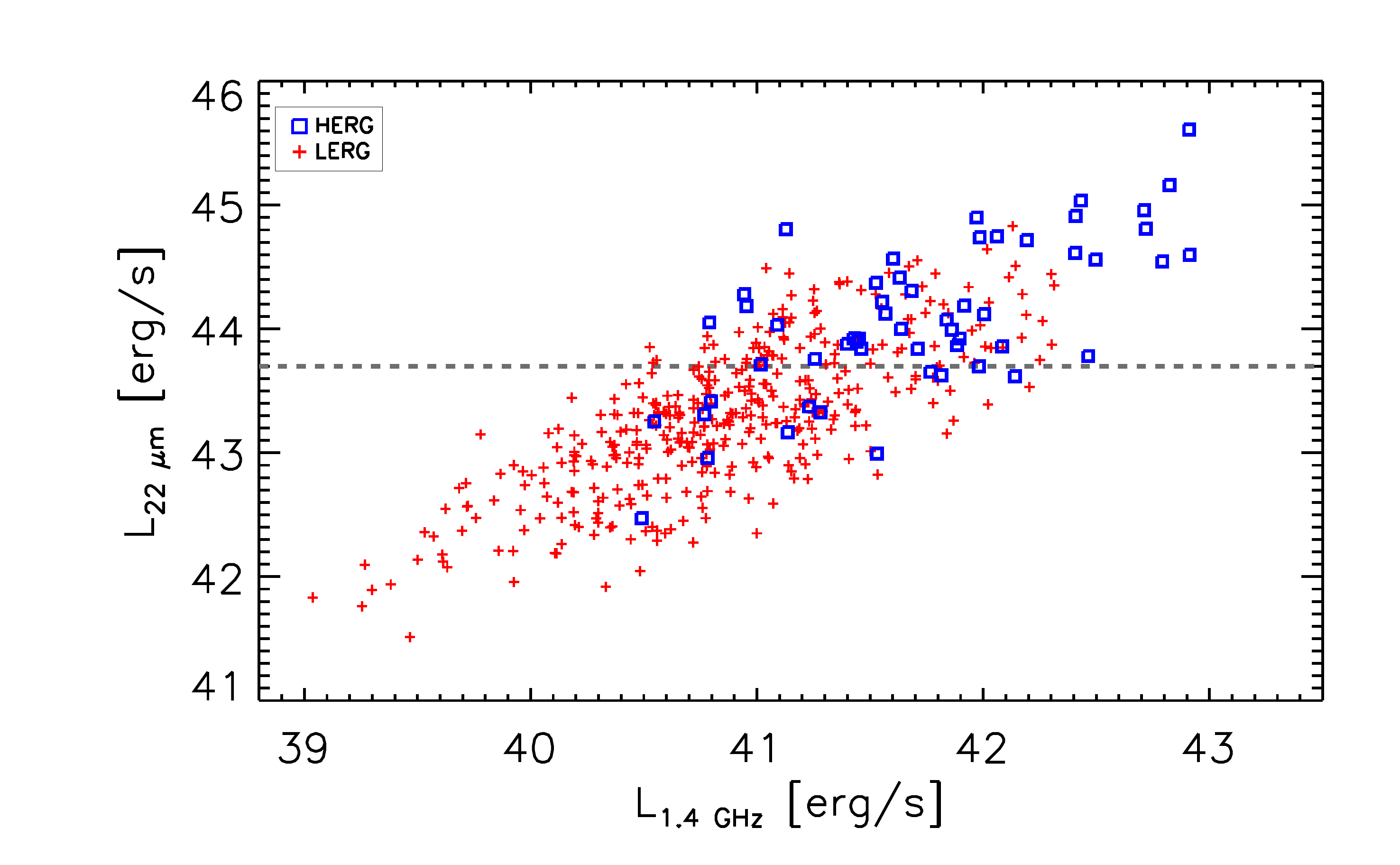}
      \caption{Mid-IR luminosity ($L_{22\mu m}$, in erg/s) versus 
      radio luminosity ($L_{\rm 1.4\,GHz}$, in erg/s), for LERGs (red plus symbols) 
      and HERGs (blue squares). The dashed line corresponds to the 
      critical mid-IR luminosity of $5\times10^{43}$~erg/s defined by Gurkan et al.~(2014). 
      See Section 3.5.2 for details. 
}
  \label{lum_wise}
\end{figure}

Gurkan et al.~(2014) analysed the IR properties of a large sample (346) of radio-loud 
AGN including both nearby and high redshift sources ($0.003 < z < 3.4$). They found that 
their sample could be clearly divided into HERGs and LERGs in the mid-IR (22$\mu m$) radio-luminosity 
plane, with a critical mid-IR luminosity of $L_{22\mu m}\sim 5\times10^{43}$~erg~s$^{-1}$. 
If we divide our sample ($z<0.7$) based on this mid-IR luminosity, we find that 75\% of LERGs 
have $L_{22\mu m}$ below $5\times10^{43}$~erg~s$^{-1}$ and 78\% of HERGs have mid-IR luminosities 
above this value. Figure~\ref{lum_wise} shows there is a large scatter of 1 to 2 orders of magnitude about this 
dividing luminosity, however, it may be a potentially useful diagnostic tool in the absence of optical spectroscopy. 

\subsubsection{Eddington-Scaled Accretion Rate}

Recent studies have investigated the idea that a fundamental switch in the type of 
radio-loud AGN activity occurs at an accretion rate of $\sim$1\% of the 
Eddington accretion rate (e.g.~Mingo et al.~2014). At Eddington ratios greater 
than 1\%, the accretion disk structure is expected to represent the classical picture of an 
AGN in unified models with a geometrically-thin accretion disk 
(e.g. Urry \& Padovani 1995), while at accretion rates much less than 
1\% of Eddington there is a geometrically-thick radiatively-inefficient accretion flow (e.g.~Narayan \& Yi 1995). 
 
In order to estimate the Eddington-scaled total accretion rate for the polarized HERGs 
and LERGs, we follow the approach adopted by BH12. They combined the optical bolometric 
luminosity, estimated from the OIII emission line ($L_{\rm bol,OIII}$), with the total mechanical 
luminosity of the radio jet ($L_{\rm mech}$), normalised to the Eddington luminosity ($L_{\rm Edd}$),  
to obtain the Eddington-scaled total accretion rate as $(L_{\rm mech}+L_{\rm bol,OIII})/L_{\rm Edd}$.

To derive a mean bolometric correction for the SDSS [OIII]$_{5007}$ emission line luminosity, 
Heckman et al.~(2004) used a multi-wavelength analysis of a large sample of powerful Type 1 AGN to find 
\begin{equation}
L_{\rm bol,OIII} \approx 3500 L_{\rm OIII},
\end{equation} 
with a scatter in $L_{\rm bol,OIII}$ of $\pm0.38$~dex. 
The [OIII]$_{5007}$ emission line is used because it is the strongest emission line in the majority 
of SDSS AGNs, and has minimal contamination from star-formation.
Since this relation was derived mainly at low redshift ($z\sim0.1$) its applicability up to redshifts 
of $z\sim0.5$ is questionable. However, Bernardi et al.~(2003) found relatively little luminosity evolution in 
a study of 9000 early-type galaxies from the SDSS, in the redshift range $0 < z < 0.3$. 
Furthermore, the metallicity evolution in star-forming galaxies from $z\sim0.1$ to $z\sim0.4$ is 
$\sim0.1$~dex (Lara-Lopez et al.~2009). The metallicity evolution of early-type galaxies would 
not be expected to change more dramatically than this over our redshift range. Thus, considering 
the quoted uncertainty for $L_{\rm bol,OIII}$ of $\pm0.38$~dex, we do not expect Eqn.~1 to be 
strongly affected by changing abundances in our redshift range up to $z\sim0.5$. However, we will 
also check the significance of our results after excluding sources with $z > 0.2$. 

To estimate the jet mechanical luminosity from the 1.4~GHz radio luminosity, we use the relation 
of Cavagnolo et al.~(2010), 
\begin{equation}
L_{\rm mech} = 7.3\times10^{36}(L_{1.4~{\rm GHz}}/10^{24}{\rm~W~Hz}^{-1})^{0.7}.
\end{equation}
This scaling relation between the jet mechanical luminosity ($L_{\rm mech}$) and the 
1.4~GHz radio luminosity is derived from X-ray cavity measurements in mainly low power radio 
galaxies with radio luminosities less than $10^{25}{\rm~W~Hz}^{-1}$ (e.g.~Birzan et al.~2008). 
The large scatter in the data ($\sim$0.8~dex) means that there is considerable uncertainty in the slope 
of this relation, that makes its application to high-power radio sources questionable, as well as 
potentially introducing a systematic error in $L_{\rm mech}$ between the HERG and LERG samples 
because of their different luminosity distributions. 
However, Godfrey \& Shabala (2013) used an independent method for determining the total power of 
sources with radio luminosities greater than $10^{25}{\rm~W~Hz}^{-1}$, specifically FR2-type sources. 
Despite the significantly different radiative-efficiencies of FR2 and FR1 type radio sources, Godfrey \& Shabala (2013) 
surprisingly found very good agreement with the Cavagnolo et al.~(2010) relation up to $10^{28}{\rm~W~Hz}^{-1}$. 
While this agreement is encouraging, the applicability of both the same normalisation and slope of 
Eqn.~2 across such a wide range in radio luminosity remains uncertain, with further work on a larger 
number of sources across the full radio-jet luminosity range required.

In order to calculate the Eddington luminosity,
\begin{equation}
L_{\rm Edd} = 1.3\times10^{31} M_{\rm BH}/M_{\odot}~{\rm W}, 
\end{equation}
for each of our sources, we require an estimate of the black hole mass ($M_{\rm BH}$).
Using the velocity dispersion of the host galaxy ($\sigma_*$) taken from the SDSS spectrum, $M_{\rm BH}$ 
can be estimated for a large fraction of our sample through the $M$--$\sigma_*$ relation (e.g.~Tremaine et al.~2002). 
This relation is based on the well determined dependence of the stellar velocity dispersion of a 
galaxy bulge on its black hole mass, described by 
\begin{equation} 
\log(M_{\rm BH}/M_{\odot})= 8.13 + 4.02\log(\sigma_*/200~{\rm km~s}^{-1}).
\end{equation} 
The SDSS stellar velocity dispersion estimates are obtained by fitting a spectral template 
across the rest-frame wavelength range 400-700~nm, with emission lines regions explicitly masked out. 
The velocity dispersion is only estimated for $z < 0.4$ and we have excluded any values above 
420~km~s$^{-1}$ and below 70~km~s$^{-1}$, since they cannot be reliably measured from the SDSS spectra. 
We also excluded any spectra with a median per-pixel signal-to-noise of less than 10.\footnote{See http://classic.sdss.org/dr2/algorithms/veldisp.html for details.}
This leads to 546 reliable velocity dispersion values (22 HERGs and 524 LERGs) with a median 
redshift of 0.15 and with only four sources with $z > 0.3$. Thus, we find reliable velocity dispersion 
measurements for 40\% of polarized HERGs and $\sim$71\% of polarized LERGs. The lower fraction 
for HERGs is mainly due to a larger fraction of sources at higher redshift and with lower signal-to-noise. 

The stellar velocity dispersion as measured from the fixed 3" fiber aperture of the SDSS does 
not provide an accurate estimate of $M_{\rm BH}$ in disk-dominated galaxies because stars 
from the extended regions of the galaxy will be included. 
All the polarized HERGs and LERGs in our sample have a concentration-index ($C=R90/R50$) greater than 2.6, where the 
parameter $R90$($R50$) represents the radius that encloses 90\%(50\%) of the host galaxy optical light. 
A concentration-index greater than 2.6 means that they are bulge dominated systems (Shimasaku et al. 2001), 
as expected for the typical early-type hosts of radio-loud AGN. This means that the velocity dispersion 
as measured from the fixed 3" SDSS fibers ($\sim$8~kpc at $z=0.15$) is representative of the bulge 
in the majority of cases. 
However, at redshifts, $z>0.2$, the inclusion of the majority of the host galaxy light may affect the velocity dispersion 
measurement and the $M_{\rm BH}$ estimate in a systematic manner.
At redshifts, $z<0.1$ and $0.1<z<0.2$, we found median velocity dispersions of 240~km~s$^{-1}$ 
and 241~km~s$^{-1}$, respectively, while from $0.2<z<0.3$, the median velocity dispersion increases 
to 257~km~s$^{-1}$. This means that Eddington-scaled accretion rates for sources with $z>0.2$ are 
typically $\sim$1.6 times systematically larger than sources at $z<0.2$. 

\begin{figure}
\centering
\vspace{-0.5cm}
    \includegraphics[width=8.5cm]{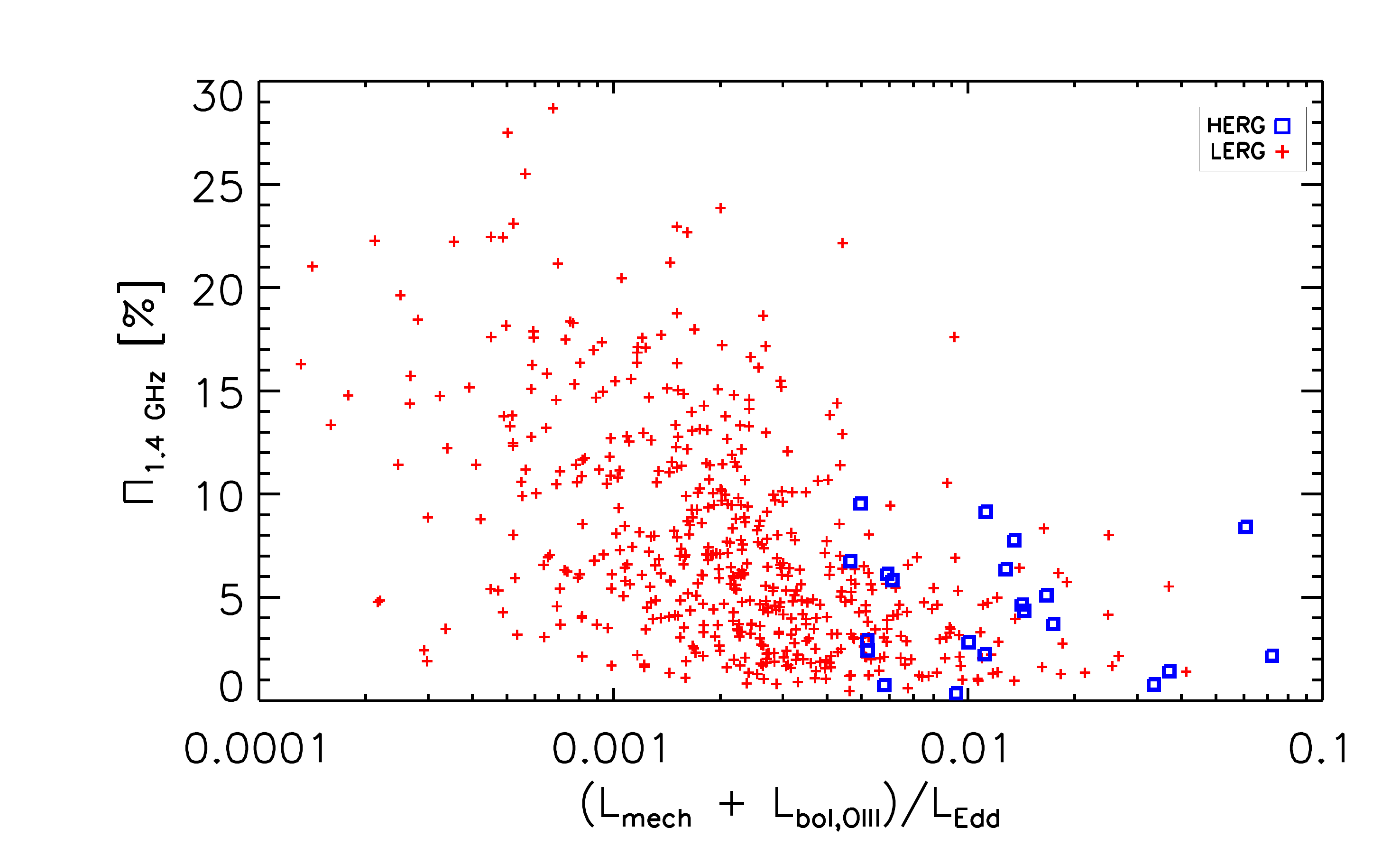}
      \caption{Integrated degree of polarization at 1.4~GHz ($\Pi_{\rm 1.4\,GHZ}$) 
      versus the total Eddington-scaled accretion rate, $(L_{\rm mech}+L_{\rm bol,OIII})/L_{\rm Edd}$, 
      for HERGs (blue squares) and LERGs (red plus symbols). See Section 3.5.3 for details. 
}
  \label{edd_ratio}
\end{figure}

In Figure~\ref{edd_ratio}, we plot $\Pi_{\rm 1.4\,GHz}$ versus the Eddington-scaled total accretion rate, 
corresponding to $(L_{\rm mech}+L_{\rm bol,OIII})/L_{\rm Edd}$, for 22 polarised HERGs and 524 polarized LERGs.
We see that the HERGs have systematically higher accretion rates 
($\sim 0.5 - 10\%$) than the majority of the LERGs that have accretion rates $< 1\%$.
In addition to this, it is clear that the more weakly accreting LERGs are more 
highly polarized in comparison to the HERGs and the strongly accreting LERGs. 
As noted by Mingo et al.~(2014), the LERGs with Eddington ratios greater than 
$\sim$1\% may be mis-identified as radiatively inefficient sources and/or their radio 
luminosity is boosted by residing in dense environments. Thus, it is interesting 
to test the significance of the difference in $\Pi_{\rm 1.4\,GHz}$ between sources 
separated purely by their total Eddington ratios. 
A two-sided KS test indicates a highly significant difference in $\Pi_{\rm 1.4\,GHz}$ 
for sources separated at a total accretion rate of 0.5\% of Eddington (p-value of $1.9\times10^{-15}$, 
corresponding to $\sim$$8\sigma$ for a normally-distributed process). 
We separated the sample at 0.5\% of Eddington, instead of 1\% of Eddington, in order 
to include all the HERG sources and better match the numbers in the two bins. 
If we exclude sources with $z>0.2$, due to their systematically higher Eddington luminosities, 
the KS test gives a p-value of $8.4\times10^{-9}$ ($\sim$$6\sigma$). 

Mingo et al.~(2014) found that the mid-IR bolometric luminosity ($L_{\rm bol,IR}$) 
provided a more reliable estimate 
of the accretion rate in HERGs. Thus, we use this as a consistency check against 
the accretion rates estimated using the [OIII] emission line. 
From the WISE 22$\mu m$ flux density, we can estimate 
$L_{\rm bol,IR}$ from the scaling relation of Runnoe et al.~(2012), where 
$\log(L_{\rm bol,IR})=15.035+0.688\,\log(\lambda L_{\lambda})$. 
While we find that $L_{\rm bol,IR}$ can be up to an order of magnitude larger 
than $L_{\rm bol,OIII}$ for the HERGs, the qualitative difference when $L_{\rm bol,IR}$ is used to estimate 
the total accretion rate instead of $L_{\rm bol,OIII}$ does not result in a significantly 
cleaner separation in the Eddington ratios between polarized HERGs and LERGs.  
This also does not hugely effect the significance of the difference in $\Pi_{\rm 1.4\,GHz}$ 
for sources separated at an Eddington ratio of 0.5\% (p-value of $1.3\times10^{-8}$, $\sim$$5.7\sigma$). 

The bolometric luminosity correction in Eqn.~1 is for the [OIII] flux uncorrected for dust 
extinction. This correction can be important because while the [OIII] line emission region lies 
outside the dusty torus of the AGN (i.e.~in the Narrow Line Region), it may still suffer significant 
amounts of extinction due to interstellar dust in the host galaxy. To investigate this potential error 
in $L_{\rm bol,OIII}$, we corrected for the effect of 
dust extinction on the [OIII] flux using the Balmer decrement (i.e.~the ratio of the fluxes of the 
H$\alpha$ and H$\beta$ narrow emission lines, see~Lara-L\'opez et al.~(2009) for details). 
The median size of the dust correction to the [OIII] luminosity for our sample is $\sim$0.2. 
Kauffmann \& Heckman (2009) used X-ray, optical and mid-infrared spectra of a complete 
flux-limited sample of SDSS type 2 AGN to derive a mean bolometric correction of 
600 for the extinction-corrected [OIII] luminosity. Using this bolometric correction to the 
extinction-corrected [OIII] luminosity, instead of Eqn.~1, decreases the significance in 
$\Pi_{\rm 1.4\,GHz}$ between low and high Eddington ratios to $\sim$$4.8\sigma$. 
However, this decrease in significance is unsurprising since we lose a large fraction 
of our sources because we can only obtain the extinction correction for $\sim$50\% of sources 
due to the absence of H$\alpha$ and/or H$\beta$ measurements. 

While the different bolometric correction factors produce slightly different results and their 
application at redshifts $z>0.2$ is questionable, the main result of this section, that $\Pi_{\rm 1.4\,GHz}$ 
can achieve higher values for sources with lower total Eddington accretion rates, 
remains valid, for a separation at $\sim$0.5\% of Eddington. 

\section{Discussion}

Our main observational result is the discovery of a difference in the integrated 
degree of polarization at 1.4~GHz ($\Pi_{\rm 1.4\,GHz}$) between radio-loud radiative-mode 
AGN (i.e.~HERGs and radio-loud QSOs) and jet-mode AGN (i.e.~LERGs), with the 
weakly accreting sources extending to higher values of $\Pi_{\rm 1.4\,GHz}$. 
We now discuss the potential significance of these results in linking the large-scale host galaxy
environment with the accretion state of the supermassive black hole and the production 
of powerful radio-loud jets. 

\subsection{Environments of HERGs and LERGs}
At low redshift ($z<0.4$), HERGs are found mainly in low galaxy-density 
environments (i.e.~groups), while LERGs are found in a wide range of 
environments from poor to rich groups and clusters (e.g.~Best 2004, Hardcastle 2004). 
They also differ in their host galaxy types, with HERGs mainly residing in galaxies 
with significant on-going star formation implying an abundance of cold gas, with 
LERGs mainly hosted by massive elliptical galaxies with hot X-ray halos and a minimal 
supply of cold gas. This has led to the development of models claiming 
that the two different types of AGN result from their presence in different environments. 

In particular, Hardcastle et al.~(2007) proposed that the difference between HERGs and 
LERGs is due to the \emph{source} of the accreting gas: Bondi accretion of the hot, X-ray 
gas in ellipticals for LERGs, and cold gas accretion from disk-galaxies required for HERGs. 
Alternatively, the type of radio-loud AGN activity may instead be triggered by a 
fundamental change in the accretion disk structure at an Eddington-scaled accretion rate 
of $\sim$1\% (e.g.~Mingo et al.~2014), with a radiatively-inefficient accretion flow below, 
and a radiatively-efficent thin disk above, $\sim$1\% of Eddington. However, these two scenarios 
are not mutually exclusive and both could potentially act together. 
It remains unclear as to what the exact nature of the relationship is between the large scale 
environmental properties of AGN and the type of accretion onto their supermassive black holes. 

In Section 3.5, we have shown that our sample of radio-polarized AGN display 
similar host galaxy properties and accretion rates as the full sample in BH12 and elsewhere 
(i.e.~the host galaxies of our sample of HERGs typically have higher rates of star formation, 
lower black hole mass and higher accretion rates than the LERGs). 
Thus, we can use our sample of polarized sources to reliably infer some of the key 
properties of the HERG and LERG populations in general.

The polarized sources in our sample are in less dense 
environments (in terms of the galaxy number density within 0.5 Mpc) than unpolarised sources. 
Furthermore, it is the most underdense environments of the polarized LERG sources that typically  
have higher values of $\Pi_{\rm 1.4\,GHz}$ (Section 3.4.3). With a similar approach, Shi et al.~(2010) studied 
the most highly polarized sources in the NVSS and found a median 
galaxy count within a 1 Mpc radius of $7\pm3$ (for only eight sources) compared to $9\pm7$ 
for a much larger low-polarization control sample. Their result is not inconsistent with our 
findings, and is suggestive of an influence of galaxy number density on $\Pi_{\rm 1.4\,GHz}$. 
Overall, galaxy number density estimates seem to be crude but statistically useful probes of the 
environment densities of radio sources in relation to whether they are in a poor or rich group, or cluster. 

In considering the radio source environment as a key variable in our analysis, we also need to 
recognise that bent or disturbed morphology radio sources are often found in dense environments 
(e.g.~Owen \& Rudnick 1976). In fact, radio sources with bent morphologies are used as a way 
to find candidate galaxy clusters (e.g.~Wing \& Blanton 2011). In Section 3.3.2, we showed that 
bent radio sources have significantly lower values of $\Pi_{\rm 1.4\,GHz}$ than straight ones. 
We also found that $\Pi_{\rm 1.4\,GHz}$ for LERG sources was weakly correlated with the total intensity spectral 
index (Section 3.4.4), providing additional evidence that the LERGs with low values of 
$\Pi_{\rm 1.4\,GHz}$ are preferentially located in denser environments.  
This reduction in $\Pi_{\rm 1.4\,GHz}$ could be due to a combination of polarization angle cancellation, due 
to asymmetric jet structure, as well as the large amounts of external Faraday rotation expected in 
dense environments, that would cause strong depolarization of the radio emission. 

These results are generally consistent with Best~(2004) who found, for $z < 0.1$, a strong correlation 
between radio luminosity and environment richness for LERGs, but not for HERGs. 
Similarly, Ineson et al.~(2013) found, for radio sources around $z\sim0.4$, a weak 
correlation between radio luminosity and cluster richness for LERGs, but no correlation 
for HERGs. However, if the environment of radio sources does indeed strongly affect the integrated 
degree of polarization, the question still remains as to how exactly this is achieved. 
Thus, a more sensitive probe of the immediate environment of radio sources is desirable for probing the 
link between galaxy environment, black hole accretion and powerful jet production in more detail.
One of the most sensitive probes of the local environment of radio sources is through Faraday 
rotation and its associated depolarisation. In Sections 3.4.1 \& 3.4.2, we showed that the 
currently available Faraday rotation and depolarisation data for our sources is not sufficient to probe 
the local source environment in detail, however, it is clear that the majority of depolarisation occurs 
local to the source (e.g.~Banfield et al. 2014, Farnes et al.~2014).

From our current investigation, we expect that the LERGs with high values of $\Pi_{\rm 1.4\,GHz}$ 
are suffering less depolarization in less dense environments than the LERGs with low values 
of $\Pi_{\rm 1.4\,GHz}$. This is supported by our observation that sources with straight 
jets are much more likely than bent jets to have $\Pi_{\rm 1.4\,GHz}>15\%$ (Section 3.3.2), 
and the environment density estimates discussed above indicate that the more highly polarized 
sources are more likely to exist in less dense environments. 
From simulations, the amount of depolarization is expected to decrease with increasing lobe 
size (e.g.~Hardcastle \& Krause 2014). However, we find that the sources with 
$\Pi_{\rm 1.4\,GHz}>15\%$ have smaller linear sizes (Section 3.3.3), possibly indicating 
that these LERGs also have intrinsically weaker jets. 
While the differences between FR1 and FR2 radio morphologies are generally attributed to a 
combination of jet power and how they interact with their ambient environments (i.e.~the 
less powerful jets of FR1s are gradually decelerated by entrainment while 
the more powerful FR2s are not), our results in Section 3.3.1 suggest that the FR morphology 
is not the primary driver causing the observed difference in $\Pi_{\rm 1.4\,GHz}$ of the LERGs 
and HERGs. 

Our claim that the density of the ambient environment is the key variable for the 
spread in $\Pi_{\rm 1.4\,GHz}$ for LERGs does not explain the fact that HERGs are 
limited to $\Pi_{\rm 1.4\,GHz}<15\%$, since they are expected to exist in less dense 
environments, on average, than LERGs. A possible remedy to this could be 
related to the much greater radiative output of the HERG central engine. The high 
radiative output is likely photo-ionising large amounts of the host galaxy interstellar 
medium and increasing the amount of `Faraday-active' material (i.e.~the amount 
of magnetoionic material contributing to the total Faraday rotation as seen by the jet). 
The larger amount of magnetoionic material would also mean larger variations in 
Faraday rotation causing greater amounts of depolarization in HERGs, on average, 
than in LERGs. 
There is a well studied correlation between the AGN accretion disk luminosity and 
the amount of the host galaxy that is ionised (e.g.~Netzer et al.~2006). In fact, in some cases 
there is evidence of saturation of the ionisation radius, indicating that the AGN has 
potentially ionised the entire galaxy (e.g.~Curran \& Whiting 2010, Hainline et al.~2014). 
However, the central AGN need not be responsible for the ionisation of the large scale host 
galaxy environment in all cases. Extended emission line regions (EELRs), on scales of 
$\sim$100~kpc, are often found in sources with strong nuclear emission lines and are 
coincident with regions of high depolarization in the lobes of radio galaxies (e.g.~Pedelty et al.~1989). 
There is also statistical evidence that these EELRs are aligned with the radio source axis 
and have similar extents to the radio lobes (McCarthy et al.~1987, Baum \& Heckman 1989). 
This suggests that the radio jet may also be responsible for the ionisation of its 
immediate environment through shocks, and shock-induced star formation, as well as entrainment. 
Interestingly, from a Faraday rotation measure study of 26 radio-loud AGN ($0.3 < z < 1.3$), 
Goodlet \& Kaiser~(2005) found that the difference in the RM of each lobe and the 
dispersion in RM both correlate with redshift, suggesting more dense magnetoionic 
environments at higher redshift. Since all their high-redshift sources are HERGs and 
their low-redshift sources are dominated by LERGs, this lends support to the argument 
that HERGs maximum integrated polarization may be limited due to large 
amounts of Faraday depolarization local to the source.  
Once higher precision Faraday rotation measure and depolarisation data are obtained 
for a large number of HERGs and LERGs, it would then be interesting to consider other 
factors, in addition to the magnetoionic environment, such as the frequency of gas-rich 
mergers as a function of redshift and environment, 
the star formation histories of the host galaxies and their X-ray properties, for example.

Banfield et al.~(2014) also studied the 1.4~GHz polarization properties of a large 
sample of radio galaxies and radio-loud QSOs taken from the catalog presented in Hammond et al.~(2012).
Banfield et al.~(2014) discussed the difference in the degree of polarization between radio 
galaxies and radio-loud QSOs as a result of a cosmic evolution in the space density 
of quiescent galaxies (i.e.~the typical hosts of LERGs). They also found no significant difference in the RM distribution 
of the two types of sources but concluded that most of the extragalactic RM must 
originate close to the source. Such studies have important implications for predicting 
the number density of polarized sources that will be detected in deep polarization 
surveys with the SKA and its precursors/pathfinders. The integrated degree of polarization 
of radio-loud AGN at 1.4~GHz has been claimed to be anti-correlated with the total radio flux density 
(e.g.~Mesa et al.~2002, Tucci et al.~2004, Taylor et al.~2007, Subrahmanyan et al.~2010, Stil et al.~2014), 
and although this relation is disputed (Hales et al.~2014), it was claimed that 
it is the intrinsically less luminous radio sources that are more highly polarized (Banfield et al.~2011, 2014). 
Preliminary efforts were made to explain the anti-correlation based on a change in the 
dominant population of polarized radio sources from FR2s to FR1s towards lower 
flux densities (O'Sullivan et al. 2008, Stil et al.~2014). However, the results of this paper 
strongly suggest this is not the case. 
We also find a weak but significant anti-correlation between the 1.4~GHz radio luminosity 
and $\Pi_{\rm 1.4\,GHz}$ (Section 3.2) and instead claim that it is most likely driven by the 
difference in $\Pi_{\rm 1.4\,GHz}$ between radio-loud radiative-mode AGN, that dominate at high 
luminosities, and the jet-mode AGN, that dominate at lower luminosities. 

\subsection{Jet production efficiency}
After liking the large scale magnetoionic environment of 
radio sources to their host galaxy accretion states (i.e.~HERGs and LERGs), 
it is interesting to discuss the potential implications of the 
accretion of magnetised gas from large distances in enabling the production of 
powerful, radio-loud jets. 

The existence of large amounts of poloidal magnetic flux close to a supermassive 
black hole is synonymous with jet production (e.g.~McKinney \& Blandford 2009, Porth et al.~2011, Fendt et al.~2014) 
and also with the extraction of the rotational energy of the black hole in the so called magnetically-arrested 
accretion disk regime (MAD; Igumenshchev et al.~2003). In MAD systems, it is expected that the 
mass accretion rate and jet magnetic flux is strongly correlated (Tchekovskoy et al.~2011), 
which is supported by recent observational results (Zamaninasab et al.~2014, Zdziarski et al.~2014). 
However, these observations are of sources with radiatively-efficient accretion and 
the MAD systems have only been successfully simulated in radiatively-inefficient regimes. 
Furthermore, van Velzen et al.~(2013) have argued that the MAD model is inconsistent with 
the radiatively-efficient accretion in radio-loud QSOs, on the grounds that the spread in 
the optical-radio luminosity relation does not allow for a large range of black hole spins. 
Mocz \& Guo (2015) presented a model for interpreting MADs in both the radiatively-efficient and 
radiatively-inefficient accretion regimes, however this is based on the assumption a MAD can 
be successfully formed in the radiatively-efficient accretion regime. 

In any case, the required amounts of magnetic flux to generate such strong poloidal fields close to the 
black hole most likely needs to be advected from large distances, since the generation 
of such large magnetic flux in-situ is considered unlikely (e.g.~Begelman 2014). 
If the differences in accretion mode of HERGs and LERGs is mainly due to the large 
scale environment (i.e.~HERGs have a radiatively-efficient accretion flow due to a 
large supply of cold gas and LERGs have a radiatively-inefficient accretion flow due 
to a hot interstellar medium), then the ability of the accretion flow to form a MAD system 
and produce powerful radio-loud jets may be related to the large scale \emph{magnetised} 
environment. Thus, the study of the large scale magnetoionic environment of HERGs and 
LERGs through their polarization and Faraday rotation properties may provide important 
constraints on the likelihood of formation of a MAD. 

In Section 3.5.3, we found that the sources with low Eddington-scaled accretion rates 
($L_{\rm total}/L_{\rm Edd} < 0.5\%$) could achieve higher values of $\Pi_{\rm 1.4\,GHz}$ 
than the sources with $L_{\rm total}/L_{\rm Edd} > 0.5\%$. This could be explained 
by a more uniform jet/lobe magnetic field structure and/or less Faraday depolarization 
in the environments of sources with $L_{\rm total}/L_{\rm Edd} < 0.5\%$. 
Our analysis of the effect of radio morphology on $\Pi_{\rm 1.4\,GHz}$ suggests that 
intrinsic magnetic field differences are not the dominant effect. We also find no 
significant difference between the RMs of strongly and weakly accreting sources, but we 
claim that the currently available RM data are dominated by the RM contribution from the Milky Way and thus
insufficient to determine the true RM contribution local to the source. Higher precision 
RMs from broadband and high angular resolution radio observations should 
provide a robust test of this scenario through direct estimation of the amount of 
Faraday depolarization local to the source.

AGN that accrete in a radiatively efficient manner are 
considered to be the dominant type of AGN at high redshift and possibly responsible 
for the co-evolution of supermassive black holes and their host galaxies. 
Why only a small fraction ($\lesssim$10\%) of these radiatively-efficient AGN 
are radio-loud remains an open question. 
One proposed solution is the `magnetic flux paradigm' of Sikora \& Begelman (2013), 
who propose that the radiatively-efficient radio-loud AGN must have undergone a 
previous radiatively-inefficient accretion phase in order to build up the large amounts 
of magnetic flux required for powerful jet production (c.f.~Lubow et al.~1994). 
One way in which this could be achieved 
is by a major merger between a disk galaxy and a giant elliptical galaxy that has 
been undergoing hot accretion for some time. Such a scenario has some observational 
support in that the most luminous AGN are often triggered by mergers 
(e.g.~Ramos Almeida et al.~2012). Whether or not giant elliptical galaxies have 
sufficient amounts of coherent magnetic fields that can be dragged inwards to the 
black hole horizon is currently unclear; however, this is not inconsistent with our observations of 
highly polarized LERGs whose elliptical galaxy hosts would cause smaller amounts 
of Faraday depolarization if large-scale coherent magnetic fields were 
indeed present. 
Upcoming broadband radio polarization data from surveys, such as 
POSSUM on the ASKAP telescope (Gaensler et al.~2010) and the proposed VLA Sky Survey, 
can provide important constraints on this proposed scenario.

\section{Conclusions}

We have presented a radio polarization study of 796 radio-loud AGN 
($z<0.7$) in relation to their host galaxy properties, accretion states 
and large scale environments. 
The integrated degree of polarization at 1.4~GHz was obtained  
from the NVSS in the region that overlapped with the SDSS survey area. We used the 
SDSS optical spectroscopic classifications of Best \& Heckman~(2012), who defined a 
large sample of NVSS radio sources into high-excitation and low-excitation radio galaxies (i.e.~HERGs and LERGs). 
We find a fundamental difference in the polarization properties 
between HERGs and LERGs, where the LERGs are observed to span 
the full range of expected integrated degrees of polarization at 1.4 GHz 
($\Pi_{\rm 1.4\,GHz}$), with a maximum of $\sim30\%$, while HERGs 
(and radio-loud QSOs) are restricted to a maximum of $\Pi_{\rm 1.4\,GHz} \sim 15\%$. 
We also find that the weakly accreting LERG sources (with Eddington ratios $<0.5\%$) 
can attain higher values of $\Pi_{\rm 1.4\,GHz}$ than the strongly accreting LERGs and HERGs 
(with Eddington ratios $>0.5\%$). This may be important for attempts 
to explain the production of powerful radio-loud jets by the ability of 
AGN host galaxies to accumulate large amounts of magnetic flux close to 
the black hole (e.g.~Sikora \& Begelman 2013). 
Infrared data from WISE allowed us to determine the host galaxy type, with
LERGs mostly inhabiting early-type elliptical galaxies and HERGs typically 
with disk-dominated hosts. This is broadly consistent with detailed 
optical studies which find that the host galaxies of HERGs are typically 
of lower stellar mass and have bluer colours than LERG hosts (e.g.~Heckman \& Best 2014). 
Thus, we maintain that the results from our sample of polarized HERGs 
and LERGs are applicable to the HERG and LERG populations in general. 

The scenario that we consider to best explain the current data is one in which   
the full range of polarization from $<1\%$ to $\sim$30\% of the LERGs, can be 
explained by the full range of environments in which LERGs are observed to 
reside in, in the local universe. In high gas density environments, the jet emission structure is
more likely to be bent, causing polarization angle cancellation, while significant Faraday 
depolarization caused by the high density magnetoionic medium can help to further suppress 
the observed integrated polarization at 1.4~GHz. In poor group environments, 
the jets are more likely to remain straight and along with the lower expected amount of 
external Faraday depolarization, this leads to the highest integrated degree of polarization. 
Additionally, the LERG sources with $\Pi_{\rm 1.4\,GHz}>15\%$ have significantly 
smaller linear sizes than the other LERGs, potentially indicating that these LERGs may
also have intrinsically weaker jets. 
The restricted integrated degree of polarization in HERGs is not easily explained 
since the HERGs reside in lower galaxy density environment, on average, than 
the LERGs. However, we suggest that the high ionising luminosity of the central engine 
ionises a significant fraction of the host galaxy, increasing the amount of 
magnetoionic material contributing to the Faraday rotation local to 
the source. This would be expected to generate large amounts of Faraday depolarization 
in the HERG environment, potentially restricting the integrated degree of polarization 
at 1.4~GHz from reaching values greater than $\sim$15\%.

However, although we expect Faraday effects to dominate at 1.4 GHz, we do not 
find any direct evidence for this from the Faraday rotation measures (RMs) derived from 
the NVSS data in Taylor et al.~(2009). These RMs have been shown to be dominated 
by the magnetoionic material of the Milky Way and are relatively insensitive probes of 
the local environments of extragalactic radio sources (e.g.~Oppermann et al.~2015), 
thus, we cannot test our above hypothesis in a robust manner. 
We have also considered intrinsic magnetic field differences of the large 
scale radio morphological classes (i.e.~FR1, FR2 and FR0) as a potential origin 
of the difference in the polarization properties of the HERGs and LERGs. However, 
we do not find any conclusive evidence suggesting that the FR morphological class is the primary driver. 
To understand the relative importance of Faraday rotation and intrinsic magnetic 
field structure in determining the polarization properties of HERGs and LERGs requires 
high precision RM and polarization measurement across a wide 
range of frequencies. Upcoming broadband radio polarization surveys, such as the 
Polarization Sky Survey of the Universes Magnetism (POSSUM) on the Australian Square Kilometre 
Array (SKA) Pathfinder telescope (ASKAP) in conjunction with the proposed VLA Sky Survey will enable 
much more robust inferences on the intrinsic magnetic field and magnetoionic environment 
of HERG and LERG sources. 

\section*{Acknowledgements}

SPO'S and BMG acknowledge the 
support of the Australian Research Council through grants FS100100033 
and FL100100114, respectively. 
The authors would like to thank Jeroen Stil for advice on the polarization bias 
correction and the referee for several important comments which significantly improved the paper. 
This research has made use of NASA's Astrophysics Data System Service and 
the NASA/IPAC Extragalactic Database (NED) that is operated by the Jet 
Propulsion Laboratory, California Institute of Technology, under contract with the 
National Aeronautics and Space Administration. 
This research made use of TOPCAT, an interactive graphical viewer and editor 
for tabular data (Taylor 2005).

  \bibliography{bflux_bib}

\newpage

\begin{appendix}
\section{The brightest radio sources: $I>100~{\rm \lowercase{m}J\lowercase{y}}$}

Stil et al. (2014) showed that the median degree of polarization of all NVSS sources increases towards lower flux densities (e.g.~the presented fit to their data gives a median degree of polarization of 1.93\% at 100 mJy and 2.17\% at 10 mJy). 
Therefore, we split our sample at $I=100$~mJy in order to compare the properties of the brighter sources to the fainter sources in our sample. At $I>100$~mJy, there are 211 LERGs and 37 HERGs. We find no significant difference in $\Pi_{\rm 1.4\,GHz}$ between HERGs and LERGs for $I > 100$~mJy. The number of LERGs in our sample with $\Pi_{\rm 1.4\,GHz} > 15\%$ decreases from 71 at $I < 100$~mJy ($\sim$10\% of LERGs) to only 2 at $I > 100$~mJy ($\sim$1\% of LERGs). 

In Figure A1, we repeat Figures 4, 6, 7 \& 8 from the main text but now only include sources with $I>100$~mJy. 
Comparing Fig.~A1(a) to Fig.~4, we see that there is no strong redshift dependence on flux in our sample. 
The median redshift for LERGs changes from 0.18 to 0.16, and from 0.25 to 0.23 for HERGs. The majority (88\%) of the radio-loud QSOs have $I > 100$~mJy. 
In Fig.~A1(b) \& (c), we see that most sources with $L_{\rm 1.4\,GHz} < 10^{24}~{\rm W~Hz^{-1}}$ are removed. In fact, the median radio luminosity increases by a factor of approximately 2 for both HERGs and LERGs and there is no longer any significant evidence for a correlation between $\Pi_{\rm 1.4\,GHz}$ and $L_{\rm 1.4\,GHz}$.  
Fig.~A1(d) confirms that even for the brightest sources, the radio morphology does not show any clear split in radio luminosity into FR1 and FR2 type sources. There is no significant difference in $\Pi_{\rm 1.4\,GHz}$ between FR1, FR2 or FR0 source morphologies. 

Interestingly, the median total intensity spectral index of sources with $I > 100$~mJy is $-0.7$ while it is $-0.6$ for $I < 100$~mJy. If the spectral index is a useful probe of the environment density (as argued in Section 3.4.4), then the less-steep spectral index for sources with $I < 100$~mJy may explain why there is  a greater fraction of LERGs with $\Pi_{\rm 1.4\,GHz} > 15\%$ at these lower fluxes. This provides some support for our conclusion that a sub-population of polarized LERGs with weaker jets and smaller sizes in under dense environments can achieve $\Pi_{\rm 1.4\,GHz} > 15\%$. We have spectral index measurements for $\sim$60\% of sources with $I>100$~mJy, but for only $\sim$20\% of sources with $I<100$~mJy. The upcoming broadband radio polarization surveys will provide much better statistics in order to investigate this in more detail. 

\begin{figure}[hb]
\centering
\includegraphics[width=18.0cm]{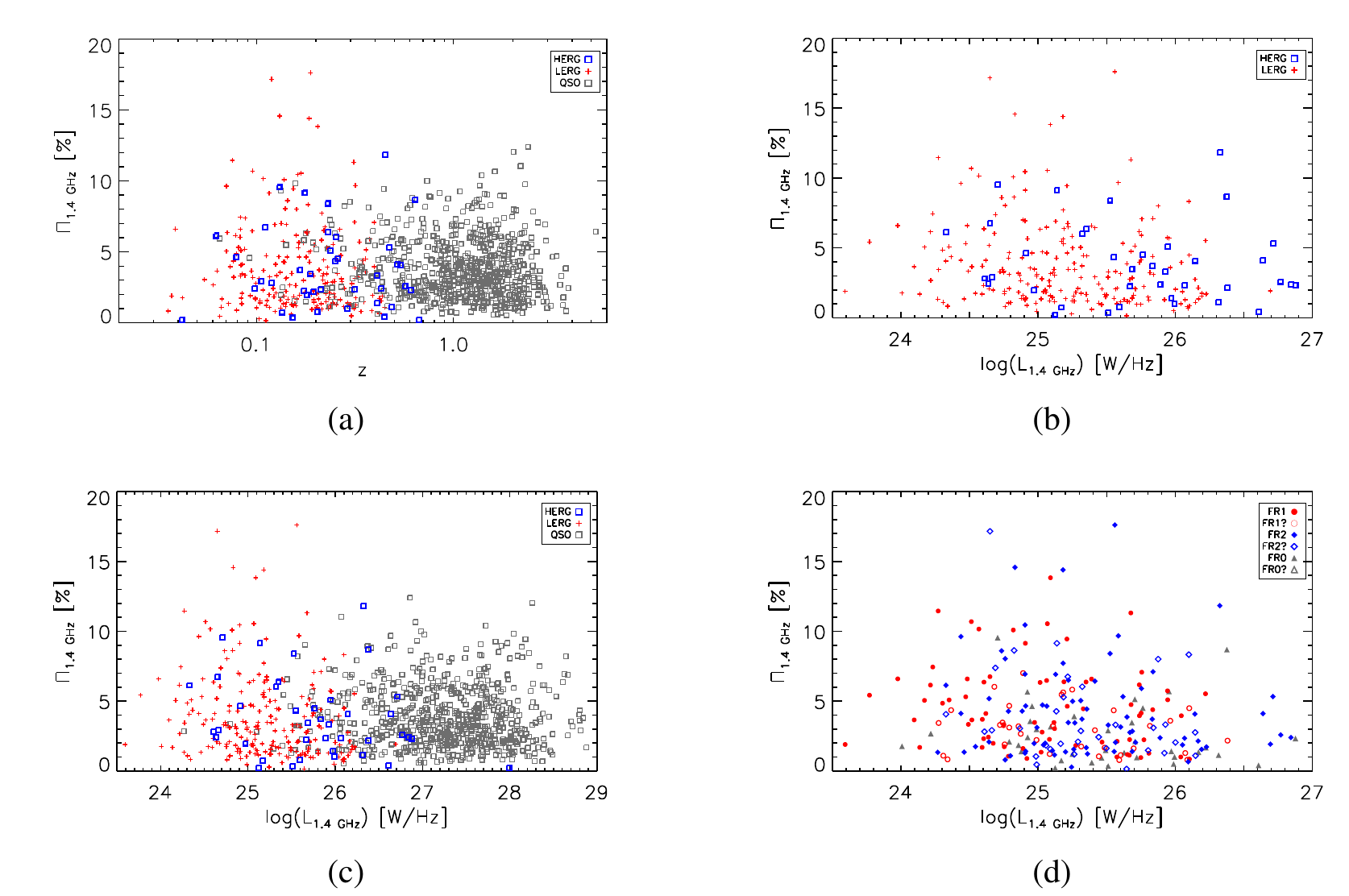}
      \caption{Figures (a), (b), (c) \& (d) correspond to Figures 4, 6, 7 \& 8 of the main text, respectively, but now only including sources with total radio flux density greater than 100~mJy.  
}
  \label{gt100mJy}
\end{figure}

\end{appendix}

\end{document}